\documentclass[fleqn,10pt,sort&compress]{wlscirep}
\usepackage[numbers]{natbib}
\usepackage[utf8]{inputenc}
\usepackage{graphicx}
\usepackage{subcaption}
\usepackage{hyperref}
\usepackage[T1]{fontenc}
\usepackage[ttdefault=true]{AnonymousPro}

\usepackage[utf8]{inputenc}
\usepackage[mathletters]{ucs}
\usepackage{textcomp}  

\usepackage{gensymb}   
\title{WaveFormer: transformer-based denoising method for gravitational-wave data}

\author[1,2,3+]{He Wang}
\author[4,+]{Yue Zhou}
\author[5,6,*]{Zhoujian Cao}
\author[3,6]{Zong-Kuan Guo}
\author[4,*]{Zhixiang Ren}

\affil[1]{International Centre for Theoretical Physics Asia-Pacific, University of Chinese Academy of Sciences, 100190, Beijing, China}
\affil[2]{Taiji Laboratory for Gravitational Wave Universe, University of Chinese Academy of Sciences, 100049, Beijing, China}
\affil[3]{CAS Key Laboratory of Theoretical Physics, Institute of Theoretical Physics, Chinese Academy of Sciences, 100190, Beijing, China}
\affil[4]{Peng Cheng Laboratory, Shenzhen, 518000, Guangdong, China}
\affil[5]{Institute of Applied Mathematics, Academy of Mathematics and Systems Science, Chinese Academy of Sciences, Beijing, 100190, China}
\affil[6]{School of Fundamental Physics and Mathematical Sciences, Hangzhou Institute for Advanced Study, UCAS, 310024, Hangzhou, China}

\affil[*]{e-mail: zjcao@bnu.edu.cn, renzhx@pcl.ac.cn}
\affil[+]{these authors contributed equally to this work}

\begin{abstract}
With the advent of gravitational-wave astronomy and the discovery of more compact binary coalescences, data quality improvement techniques are desired to handle the complex and overwhelming noise in gravitational wave (GW) observational data.
Though recent machine learning-based studies have shown promising results for data denoising, they are unable to precisely recover both the GW signal amplitude and phase.
To address such an issue, we develop a deep neural network centered workflow, WaveFormer, for significant noise suppression and signal recovery on observational data from the Laser Interferometer Gravitational-Wave Observatory (LIGO).
The WaveFormer has a science-driven architecture design with hierarchical feature extraction across a broad frequency spectrum.
As a result, the overall noise and glitch are decreased by more than one order of magnitude and the signal recovery error is roughly 1\% and 7\% for the phase and amplitude, respectively.
Moreover, on 75 reported binary black hole (BBH) events of LIGO we obtain {a significant} improvement of inverse false alarm rate.
{Our work highlights the potential of large neural networks in gravitational wave data analysis and, while primarily demonstrated on LIGO data, its adaptable design indicates promise for broader application within the International Gravitational-Wave Observatories Network (IGWN) in future observational runs.}
\end{abstract}
\begin{document}

\flushbottom
\maketitle
%
%


\section{Introduction}\label{sec1}

In September 2015, the Laser Interferometer Gravitational-Wave Observatory (LIGO) \cite{aligo} detected gravitational waves (GWs) from distant colliding black holes \cite{GW150914,PRL061102,PRX041015}, ushering in the era of GW astronomy.
Since then, dozens of merging black-hole and neutron-star binaries \cite{GWTC1,GWTC2,GWTC2.1,GWTC3} have been observed by LIGO and Virgo \cite{avirgo}.
{Currently, while some GW detection methods\cite{Burst,CBC1,CBC2} that do not need templates are emerging, searching for sources of GW still typically utilizes template-matching-based analysis \cite{MF}, which performs better in the case of stationary Gaussian noise superimposed on a precisely known signal waveform.} 
However, data collected by the {LIGO-Virgo-KAGRA} detectors contains time series of GW strains that are heavily contaminated by loud noise artifacts that are analogous to the waveforms of the actual signals, and conversely bias the analysis results of the parameters of the putative astrophysical sources \cite{glitchBBH}.
{In addressing these challenges, several non-linear noise subtraction frameworks (e.g., DeepClean \cite{DeepClean}, NonSENS \cite{NonSENS}) and glitch subtraction methods (e.g., BayesWave \cite{Bayeswave}) have been developed, aiming to improve the reliability of catalog parameter estimation}.
When a candidate signal is identified, rigorous studies are carried out to verify whether the candidate is related to instrumental causes \cite{noiseLIGO1,noiseLIGO2} or data quality issues \cite{dqligoO1,dqligoO2O3,dqvirgoO3} that could potentially impact the analysis of the candidate event with poor significance estimates, and even confute the astrophysical origin.
{In gravitational wave data analysis, noise sources can generally be categorized into two types: persistent wide band noise and short-duration noise artifacts, commonly known as glitches. The former represents noise that is continuously present at certain frequencies, while the latter refers to abrupt, transient disturbances.}
Although the noise subtraction process \cite{denoiseLIGO} can help reduce the wide band noise in the {LIGO-Virgo-KAGRA} detectors, it has no effect on the amplitude of noise artifacts that are unrelated to the addressed noise sources, making the rate of loud noise artifacts one of the primary limitations of an astrophysical search strategy's signal recovery ability \cite{dqligoO1,dqligoO2O3,dqvirgoO3}.
Therefore, it is of paramount importance to quickly assess the data quality around a candidate signal by suppressing the overall level of noise while ensuring that the astrophysical signal can be recovered before initiating the subsequent analysis that determines the presence or significance of GW candidate events.

Recent advances in artificial intelligence (AI) offer a fantastic avenue for enabling or boosting study of GW into many previously inaccessible and computationally expensive issues \cite{MLGWreview1,MLGWreview2,MLGWreview3}.
The majority of state-of-the-art machine learning algorithms, however, have trouble in dealing with real-world noise that tainted by non-stationary and non-Gaussian noise artifacts \cite{MLGWSC1}.
Alternatively, several studies \cite{ShenHuerta2019,WeiHuerta2020,ChatterjeeWen2021,Bacon2022,Murali2022,deepclean} have employed the deep learning approach to denoise the GW signals from advanced LIGO noise.
{They can be used to enhance the quality of data in phase aspects, but the amplitude recovery performance is not discussed, which is necessary and important for follow-up parameter estimation research.}
Impressive progress on the large-scale deep learning model with billions of parameters, AlphaFold2 \cite{alphafold2} for example, suggests that by scaling up data, model size, and training time in the right way, model performance \cite{bert, gpt3} on difficult low signal-to-noise ratio (SNR) denoising task might be better than that of aforementioned million-parameter-scale deep learning \cite{ShenHuerta2019,WeiHuerta2020,ChatterjeeWen2021,Bacon2022,Murali2022} methods.
This work aspires to contribute to these ongoing efforts by developing, for the first time, a billion-parameter-scale transformer-based model with highly-expressiveness background noise suppression of LIGO and {International Gravitational-Wave observatories Network (IGWN) \cite{IGWN1,IGWN2,IGWN3}} {data}.

\begin{figure*}
    \centering
    \includegraphics[width=1.0\textwidth]{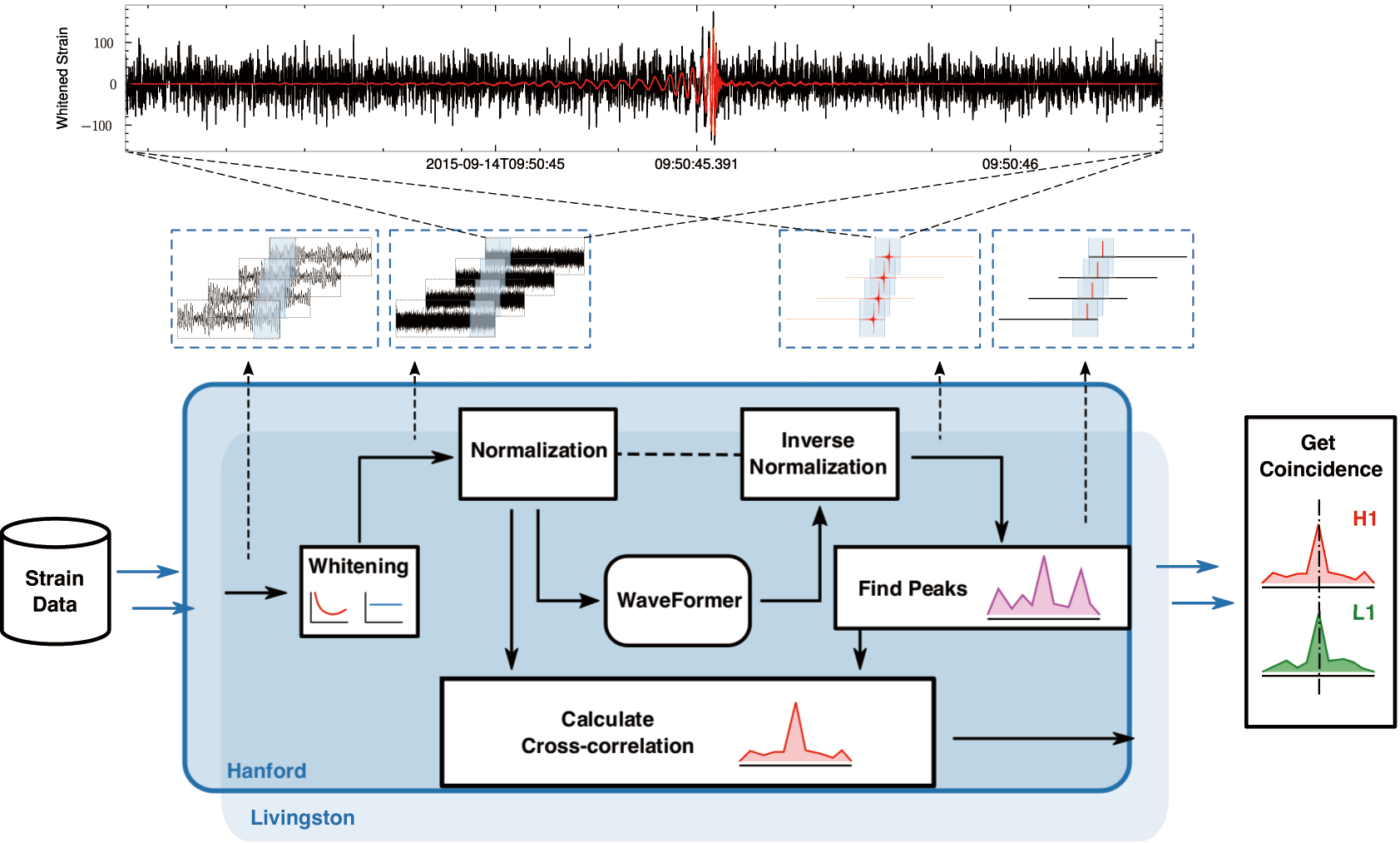}
    \caption{\textbf{GW noise suppression workflow with WaveFormer.}
    Following a preprocessing procedure including whitening and normalization, LIGO observational strain data are processed into noisy input data, which are then fed into the trained WaveFormer model to suppress the noise.
    Given the input and denoised output data of WaveFormer, a post-prcessing procedure including regular peak-finding, cross-correlation, and coincidence analysis is performed validate the denoising results.
    The waveforms on top are an example of the input and denoised output of the first BBH event, GW150914 \cite{GW150914}.
    }
    \label{fig:data}
\end{figure*}

In this article, we proposed an AI-based workflow (Figure~\ref{fig:data}) centered with WaveFormer that is designed to denoise real observational data from advanced LIGO.
Our introduced large-scale deep learning model could not only recover phase but also amplitude information of GW signals for coalescing black hole binaries.
Overall, this paper brings to the forefront some critical and proof-of-concept baselines of science-driven design for GW data analysis and aims to contribute to offline noise suppression for the observation data and GW search pipeline with the large-scale AI-based model and workflow.
At a glance, the AI-based workflow introduced in this article encompasses the following characteristics and highlights:
\begin{itemize}
    \item We proposed a large-scale AI model, WaveFormer, with science-driven innovations, including the combination of convolutional neural network and transformer for rich waveform information extraction from a wide frequency range, and masked loss for stable convergence and better denoising performance.
    \item We evaluated WaveFormer on pure noise realizations (the off-source data) and found that noise suppression was evident. The noise level percentile of the amplitude decreased from 52.5 to 0.47 and the noise amplitude spectral density (ASD) of the whole frequency range is significantly decreased from 1 to 3 orders of magnitude. With regard to GW signals that {are contaminated by different categories of glitch, the average of glitch amplitudes is 30 to 800 times smaller than before.}
    \item We further investigated WaveFormer's capacity to recover signals from observational data in terms of phase and amplitude recovery. {We achieved state-of-the-art accuracy compared with other deep learning methods\cite{WeiHuerta2020,ChatterjeeWen2021,Bacon2022,Murali2022}. On majority of the detected binary black hole (BBH) events, the phase overlaps are higher than 0.99 (1\% error)}. And no matter the circumstances, like low network SNR, we could recover the waveform amplitude with a root mean square error of $<0.53$ for matched-filtering SNR, and the typical signal recovery error is approximately 7\%.
    \item Finally, we assessed the performance of our WaveFormer-based workflow by evaluating the inverse false alarm rate (IFAR) on all reported 75 BBH events in the Gravitational-Wave Transient Catalog (GWTC), and achieved {significant} IFAR improvement, which indicated that data quality was significantly improved after noise suppression for the first time.
\end{itemize}

We showcase the trusty noise suppression performance and potential contribution to GW search of WaveFormer from multiple experiments.
The proposed AI-based workflow provides the means to enable open-source, accelerated, and deep-learning-based GW data preprocessing and the analysis has the potential to lay a solid foundation for future GW-related tasks.

\section{Methods}\label{sec4}

The overall workflow of our work is shown in Figure ~\ref{fig:data}, the raw strain data are firstly preprocessed through whitening and normalization.
After performing WaveFormer denoising, we can achieve noise suppression.
Then, inverse normalization is used to acquire the denoised observational data in the whitened domain.
Utilizing peak-finding post-processing, IFAR calculation is conducted to further evaluate our work's significance.


\subsection{Training dataset}\label{subsec7}
{Firstly, we download all the public data that released by the Gravitational Wave Open Science Center (GWOSC) {\cite{GWOSC1,GWOSC2,GWOSC3}} using CVMFS under the gwosc.osgstorage.org organization.}
{In this study, considering that waveforms from two detector are sufficient to calculate IFAR and Virgo data are not available in O1, we limited our analysis to a subset of the data, specified by 1. both Hanford and Livingston data are available; 2. data quality satisfies CBC\_CAT3 (using the GWOSC definitions) at least; 3. No hardware injections \cite{HWinj} are included; 4. No real signals: periods around known gravitational wave detections in O1, O2, and O3 are excluded.}
The final observational data that we analysed from O1, O2, O3a, and O3b consist of 48.8 days, 118.1 days, 106.7 days and 96.30 days, respectively.
These datasets are used for further significance estimates to validate our model's noise suppression performance.

Then, we generate modeled waveforms, which is frequency domain waveforms described by the IMRPhenomPv2 \cite{IMRPhenomPv21,IMRPhenomPv22,IMRPhenomPv23} model.
It covers a parameter space of BBH mergers with chirp masses of $\mathcal{M}\in[1,150]$, mass ratios of $q\in[0.125,1]$, individual masses $m_{\{1,2\}}\in[5,150]$ and spin magnitudes $a_{\{1,2\}}\in[0,0.99]$.
The time of coalescence at geocenter is constrained within an earth's rotation period on 2015-09-14 (the GPS times from 1126224017 to 1126310181.0905) to ensure generality when training.
All the other parameters are set uniformly.
In this work, we consider only short-duration BBHs.

Finally, considering a single noisy signal in the training dataset, it is a linear combination of the whitened modeled waveform and randomly sampled noise obtained from GWOSC, down-sampled at 2,048 Hz by a Butterworth filter \cite{PyCBC}.
Specifically, we extract a 32s-long noisy signal and compute its noise power spectral density.
The power spectral density is then used to whiten both the noisy signal and the modeled waveform.
{Thereafter, the SNR of the whitened noisy signal is uniformly distributed across a range from 4 to 30 by adjusting the amplitude of the signal relative to the noise level.}
We select 8.0625s-long data at center of the preprocessed 32s-long noisy signal, and normalize the standard deviation, as the input of WaveFormer.
Reason of selecting central 8.0625s-long data is to minimize impact of spectral leakage.
The ground-truth label, which is the preprocess modeled waveform, is also clipped and normalized by the corresponding standard deviation of noisy signal such that we can perform inverse normalization (Figure ~\ref{fig:data}) and recover the amplitude of the pure whiten waveforms.

{Millions of waveforms that combine noise and modeled waveform are generated for model training to overcome overfitting.}
The training dataset is augmented by randomly dropping the signals (also zeroing the labels) with 20\% probability.

\subsection{WaveFormer}\label{subsec9}

Our AI-based workflow is centered with WaveFormer, which is a deep end-to-end transformer-based pretraining model (Figure ~\ref{fig:network}) with science-driven innovations that significantly improve GW signal noise suppression performance.
The input sequence to WaveFormer is a whitened and normalized noisy signal from either Hanford or Livingston data from LIGO.
Considering an input sequence in dataset as $S_{(1,16512)}$, {which means that $S_{(1,16512)}$ is one waveform with 16512 sampling points} (8.0625-s-long waveform sampled at 2048 Hz).
Firstly, Multiple subsequences are generated through applying a fix-length window ($win\_length=0.125s\cdot2048Hz=256$) with fixed stride.
We set $stride=\frac 1 2 \cdot win\_length=128$ because a window with 50\% overlap loses minimum data information in frequency domain \cite{50overlap}.
{So, $S_{(1,16512)}$ is segmented into 128 subsequences which are stacked together and form the input ($I_{(128, 256)}$) of WaveFormer.}
One subsequence is treated as one token.
{Hence, $I_{(i,j)}$ represents $j$-th element of $i$-th token.}
The same data preprocessing method is applied to {label ($y_{(128,256)}$)} and mask.

The input data $I_{(128, 256)}$ is first embedded into dense features (DF) through embedding module as described in equation (1).
The dense features are composed of token embedding (TE), one-dimensional (1D) convolutional embedding (CE) and positional embedding (PE) .
In embedding module, input and output channel of 1D convolutional layer are both 128 and kernal size equals 3, GeLU is the activation function.
Different from one-hot vector representation for each token as in natural language processing tasks, each token of $I_{(128, 256)}$ of WaveFormer contains rich information.
Hence, CE is introduced in WaveFormer, which enrichs neighboring low-level local waveform features and high frequency signal information.
\begin{equation}
    \begin{split}
        DF &= TE+CE+PE \\
        TE &=I_{(128, 256)}W^{te} \\
        CE &= GeLU(Conv1d(I_{(128, 256)}))W^{ce} \\
        PE &= Position_{(128,128)}W^{pe}
    \end{split}
\end{equation}
Where $Position_{(128,128)}$ represents position of each token, each position is described with one-hot vector. $W^{te}, W^{ce}$ and $W^{pe}$ are embedding weights of TE, CE and PE, respectively.
Their hidden sizes are both 2048.

Then dense features are further processed through residual module as described in equation (2).
Input and output channel of two-dimensional (Conv2d) layer are both 1 and kernal size equals 7.
As illustrated in Figure ~\ref{fig:network}, the Conv2d layer is able to extract sparse spatial mid-level local features, which acts like atrous convolution in image feature extraction.
The advantage is that it can increase the receptive field and learn intermediate frequency information of signals.
\begin{equation}
    RF=Dropout(DF+GeLU(Conv2d(DF)),p=0.1)
\end{equation}

The residual feature is then fed into encoder blocks that consist of multi-head self attention module and multilayer perception (MLP).
The self attention module can extract global waveform features based on its global attention mechanism.
Compared with vanilla encoder of \cite{transformer}, some modifications are applied in WaveFormer.
Firstly, bias is removed from of MLP, because it is helpful to stabilize training process for large models \cite{palm}.
Furthermore, intermediate activation in MLP is replace with SwiGLU ($Swish(xW)\cdot xV$) because it has been shown to increase performance\cite{glu} compared with ReLU, GeLU et.al.
\begin{equation}
    \begin{split}
        Q_i,K_i,V_i &= LayerNorm(RF)(W_i^Q,W_i^K,W_i^V) \\
        Atten_i &= Dropout(Softmax(\frac {Q_iK_i^T} {\sqrt{d_k}}),p=0.1)V_i \\
        Multihead_1 &= Concat(Atten_1,...,Atten_i,...,Atten_h)W^O \\
        MSA_1 &= RF+Dropout(Multihead_1,p=0.1)
    \end{split}
\end{equation}
Where $h$ represents number of attention heads that equals 32, and hidden size of each head that equals 64. $W^O$ is the output projection of self-attention, whose hidden size equals 2048.
\begin{equation}
    Encoder_1 = Dense(SwiGLU(Dense(LayerNorm(MSA_1)))) + MSA_1
\end{equation}
Hidden sizes of inner and outer dense layer of MLP module equals 12288 and 2048, respectively.

Output of former encoder is used as input of its following encoder, and there are $m$ encoders in WaveFormer.
In our experiment, $m$ is set as 24. Finally, an output projection block and a dense layer with shared weights of token embedding layer are applied to decode output of the last encoder block, and the model output $O$ has same size as input $I_{(128, 256)}$.
\begin{equation}
    \begin{split}
    O_{proj} &= LayerNorm(GeLU(Dense(Encoder_m))) \\
    O &= O_{proj}(W^{te})^T
    \end{split}
\end{equation}

\subsection{Masked loss}\label{subsec8}
To further improve denoising performance, we propose a masked loss mechanism during WaveFormer training.
To this end, a mask is applied to calculate mean square error of WaveFormer during training.
{For each output element $O_{(i,j)}$ and its corresponding ground-truth label $y_{(i,j)}$, mean square error $e_{(i,j)}$ is defined as}:
\begin{equation}
    e_{(i,j)}=(O_{(i,j)}-y_{(i,j)})^2
\end{equation}
Masked loss is defined as:
\begin{equation}
    L=\frac {1}{i\cdot j}(\sum_{(i,j)\in mask}e_{(i,j)}+\alpha\cdot\sum_{(i,j)\notin mask}e_{(i,j)})
\end{equation}
Where $\alpha$ is a weight factor for balancing loss contribution of different elements.
In our experiment, $\alpha$ was set to $1/6$.
Compared with vanilla transformer, our introduced masked loss is in a more fine-grained form.
It can not only distinguish tokens, but also each samples within a token, which significantly accelerates convergence speed and improves training stability.
Specifically, each noisy signal has its corresponding same-shape mask.
The left and right border of the mask are calculated based on post-Newtonian theory and linear perturbation theory .
Details of mask desciption are provided in \ref{app1}.

\subsection{Injection test}\label{injtest}
To draw a more robust conclusion, we perform an injection test with real LIGO-Virgo observational data.
The noise is sampled from the first two observing runs, and then injected with a black hole waveform $h$ tuned to the desired optimal SNR \cite{OMFSNR} $\textit{SNR}_{\textit{opt}}=\sqrt{\langle h \mid h\rangle}$.
The scalar product $\langle\cdot \mid \cdot\rangle$ represents noise-weighted inner product \cite{InnerProd}.
In total, we generate a large number of injections (5000) using the same prior as in the injection test set.
All the injected templates and the denoised samples are used to calculate the matched-filtering SNR with the original injections $d$.
Within a 0.25-seconds window, we use matched-filtering SNR \cite{MFSNR} $\textit{SNR}_{\textit{mf}}=\langle d \mid h\rangle/\sqrt{\langle h \mid h\rangle}$ on the original injections for injected templates and denoised samples.

{Moreover}, we analyzed phase recovery performance on simulated compact binary coalescence signals.
To determine how well the recovered signals fit the expected waveform templates, the overlap $\mathcal{O}$ \cite{GWTC2} between them is computed as
\begin{equation}
    \mathcal{O}\left(h, h_d\right)=\frac{\left\langle h \mid h_d\right\rangle}{\sqrt{\left\langle h \mid h\right\rangle\left\langle h_d \mid h_d\right\rangle}}
\end{equation}
Where $h_d$ and $h$ are denoised waveform and whitened injection, respectively.
The scalar product $\langle\cdot \mid \cdot\rangle$ represents noise-weighted inner product \cite{InnerProd}.

\subsection{IFAR calculation}

Utilizing noise suppression results on real observational data, we further evaluate denoising performance through comparing FAR of BBH event with the public reports \cite{GWTC1,GWTC2,GWTC2.1,GWTC3,OGC1,OGC2,OGC3,OGC4}.
{Firstly, we obtain the denoised output by utilizing Waveformer.
Then, triggers are defined and identified by three steps including,
1. Find the max value in the time-series data.
2. Search nearby maximum outside 0.2s' time window of the max value in the previous step.
3. Repeat the second step until the maximum and all local-maximums (referred to as “triggers") are identified.
After finding all triggers, the following procedures are conducted to decide whether a candidate event appears,
1. By constraining triggers that exist on both two detectors, we get valid triggers.
2. We then calculate the correlation of the to-be-evaluated trigger (target trigger) between its noisy and corresponding denoised segments.
3. Through time shift, background analysis is done on other triggers around the target trigger.
Finally by counting the number of false alarm trigger pairs, we obtain the IFAR value of the target trigger, which represents the reported or candidate BBH event in this experiment.
}

For all GW events with a given IFAR observed in the data of duration $T$, we divide the data into analysis periods that allow at least 7 days (30 days for GW150914, GW151226, GW170104, GW170814, GW170809, GW170823, GW170412, GW190521\_074359, GW190707\_093326, GW200129\_065458, GW200225\_060421) of coincident data between two LIGO detectors.
The total amount of background time analyzed will equal $T_{\mathrm{obs}}=T^2 / \delta$, where $\delta$ is the time-shift interval (we set to 0.1s as same with PyCBC \cite{PyCBC})
The minimum FAR scales as $\delta / T_{\mathrm{obs}}^2$ so that approximately {7 to 30} days of coincident data are sufficient to measure FAR of 1 in $(1\sim4)\times 10^5$ years.

\section{Results and Discussion}\label{sec2}
\subsection{Science-driven deep neural network}\label{subsec1}

\begin{figure*}
    \centering
    \includegraphics[width=1.0\textwidth]{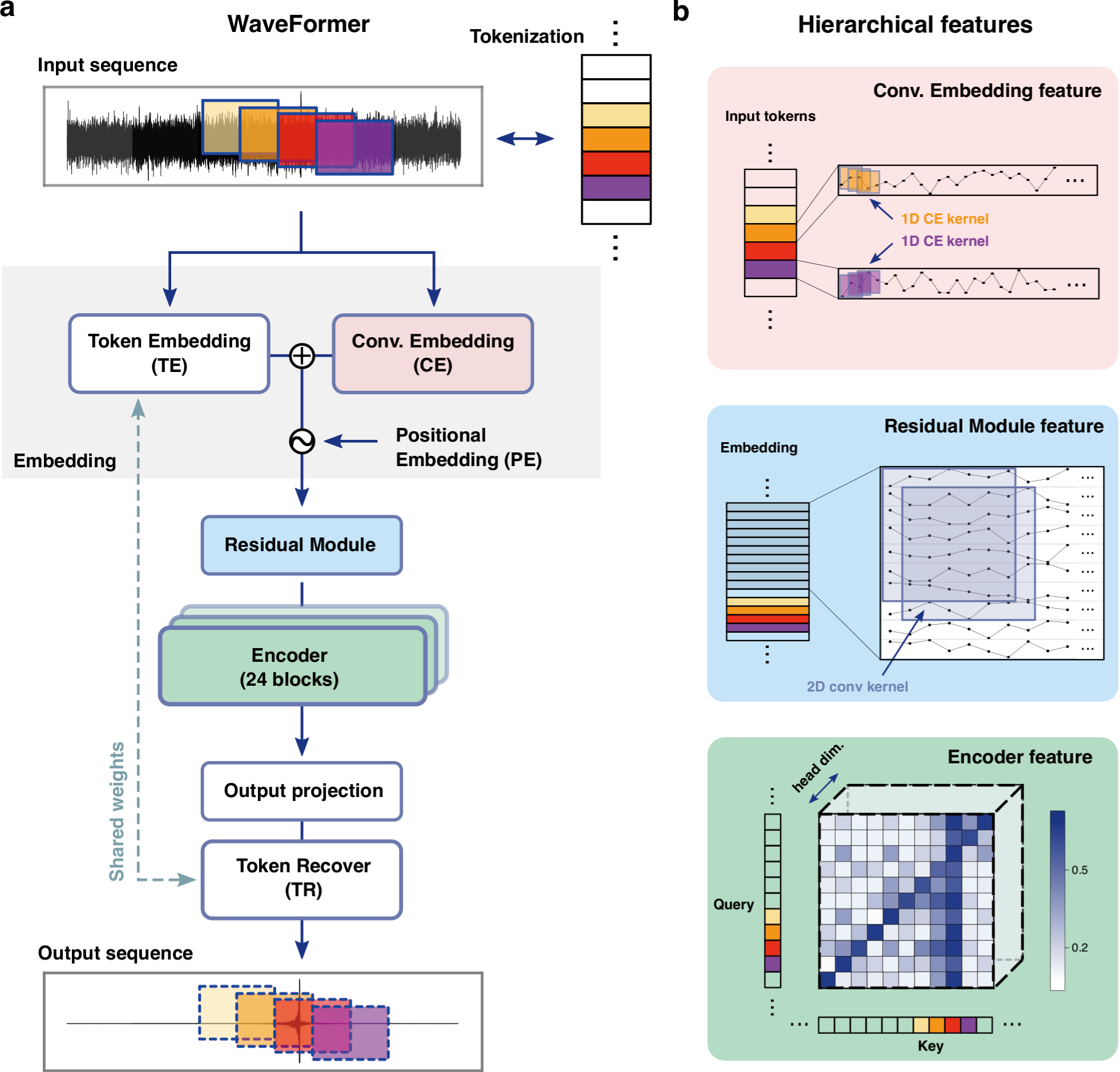}
    \caption{\textbf{Science-driven WaveFormer architecture design and hierarchical feature extractions.} a, Overview of the WaveFormer architecture. The deep neural network mainly includes embedding module, residual module, and encoders. Input and output are noisy and denoised data, respectively. b, Illustration of hierarchical feature extractions, which consists of low-level local feature (top), middle-level local feature (middle) and high-level global feature (bottom). From a science-driven perspective, they correspond to high, intermediate and low frequency information of signals. Each feature has the same background padding color as its corresponding network module in a.}
    \label{fig:network}
\end{figure*}

We design a deep learning model (Figure ~\ref{fig:network}a) that consists of stacks of transformer encoders \cite{transformer}, residual blocks \cite{residual} and embedding modules.
The model is referred to as WaveFormer.
Compared with vanilla transformer \cite{transformer}, some science-driven innovations that improve noise suppression performance are proposed in this work.
Firstly, the combination of convolutional neural networks and transformer enables our model's ability to capture generic and hierarchical features of GWs.
As depicted in Figure ~\ref{fig:network}b, convolutional embedding (CE) in the embedding module and residual module extract low-level and mid-level local features, respectively.
Encoders, on the other hand, are primarily concerned with high-level global features.
The hierarchical feature extraction mechanism is robust when applied to noise suppression tasks.
When it comes to GW signals, high-frequency information corresponds to low-level local features since they place a premium on the connections between nearby data points.
Similarly, mid-level local feature and high-level global feature correspond to intermediate- and low-frequency GW signal information, respectively, because they are more concerned with distant sampling points, such as milliseconds to seconds.
From a scientific standpoint, the comprehensive hierarchical feature extraction mechanism can process long signals and learn rich GW information of frequency domain, resulting in WaveFormer's excellent denoising performance.

Then, the dynamic mask and masked loss are introduced during network training process.
Compared to masked self-attention in vanilla transformer, our dynamic mask is in a more fine-grained manner.
We can assign {a} different mask value to each element within a token, while masked self-attention can not.
In the view of science, the importance of each sampling point for phase and amplitude recovery varies; dynamic mask can distinguish the variation and assign an appropriate mask value accordingly.
As a result, the effectiveness of GW denoising is enhanced even further.
Finally, some minor adaptions are applied to the activation and bias settings of encoders.
These adaptions have been proven to stabilize and accelerate network training and convergence.
WaveFormer is implemented based on Megatron-LM \cite{megatron} and Ray \cite{ray} framework in PyTorch \cite{pytorch}. 
For optimization, we use ADAM \cite{adam} algorithm, which works well on problems with large dataset and parameters.
Data parallel training was performed on eight NVIDIA V100 32GB GPUs and took approximately 24 hours to train for 300,000 iterations.

\subsection{Effect on realistic noise}\label{subsec3}
\begin{figure*}
    \centering
    \includegraphics[width=1.0\textwidth]{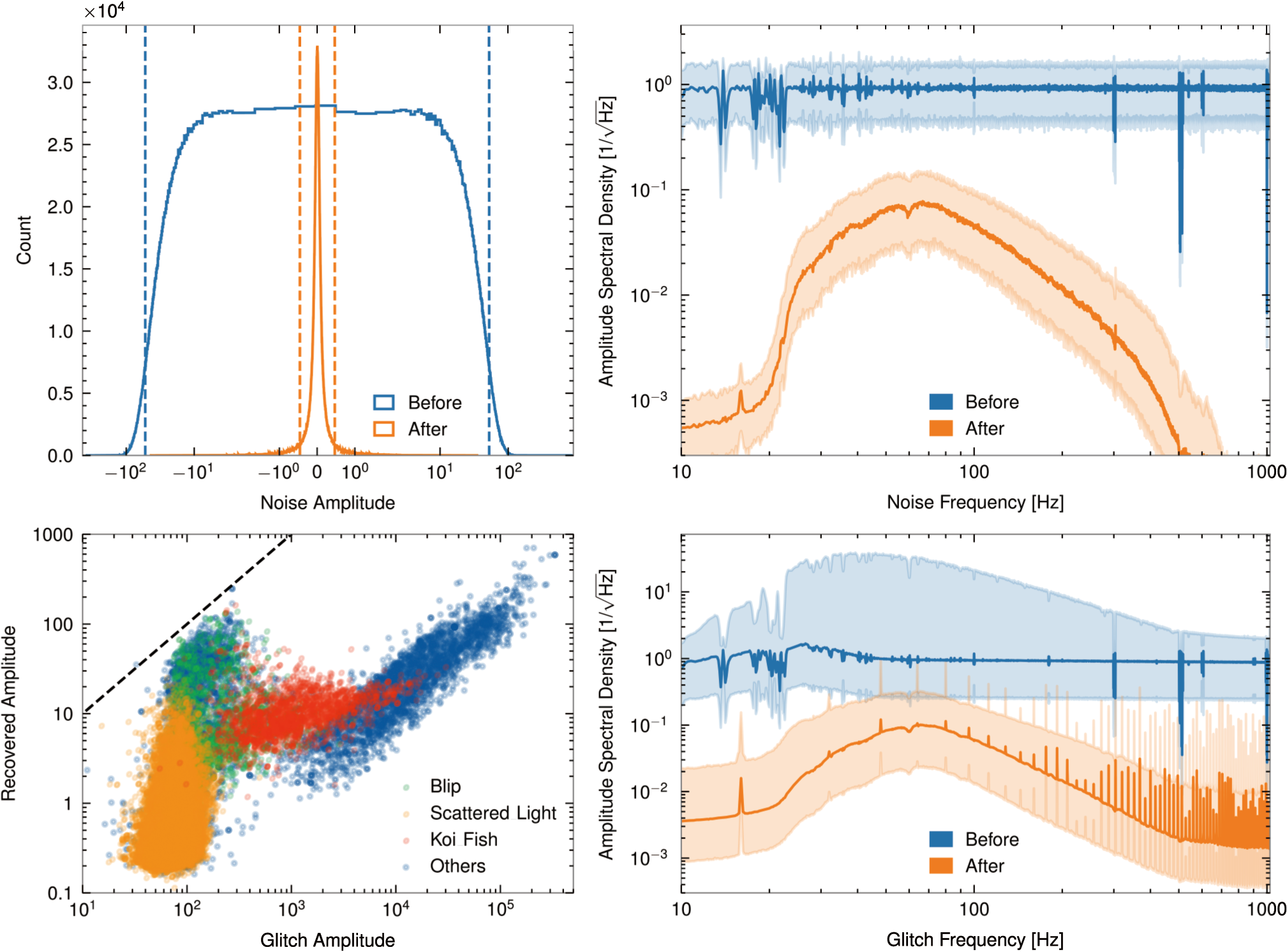}
    \caption{\textbf{Noise suppression performance on LIGO's O3b data.}
    \textbf{(Upper panels: results of pure noise)}
    The histogram distribution (left) and ASD (right) of the off-source strain data around GW200208\_130117 from the Hanford detector before and after noise suppression.
    On the histogram, the dashed lines represent the 5th and 95th percentile noise level, before=52.5 and after=0.47.
    On the spectrogram, the typical percentiles are displayed as a shaded zone surrounding the median from the 5th to the 95th percentile.
    \textbf{(Bottom panels: results of glitches)}
    (left) The amplitude of glitches (Hanford O3b) is $40\sim 400$ times lower than its original value for almost all samples.
    The dashed diagonal line means that the input glitch amplitude is equal to that of denoised glitch.
    The amplitude for the notorious blip, which is one of the limiting sources of noise for GW searches from high-mass compact binaries \cite{glitchBBH}, is 78.7 times lower than their original value.
    (right) Similar as top right panel for glitches, showing the glitch ASD is decreased by more than $1\sim3$ orders of magnitude.
    }
    \label{fig:glitch}
\end{figure*}

We present WaveFormer's noise suppression performance on real observational data by evaluating data difference after noise suppression.
{All the instances of the input data are processed with the same whitening, normalizing and denoising procedure as in our proposed workflow.}
The upper panels of Figure ~\ref{fig:glitch} showcase the 2048s-long off-source data around GW200208\_130117 in time (left) and frequency (right) domain.
The amplitude of noise level percentile is clearly compressed, reduced from 52.5 to 0.47. 
Analyzing the ASD further reveals that our WaveFormer is able to effectively eliminate narrowband and broadband spectral information while drastically decreasing the overall level of all frequency contributions.
Specifically, ASD of intermediate frequecy noise is 10 times lower after noise suppression, while ASD of low-frequency and high-frequency noise is approximately 1000 times lower than before.

Furthermore, we investigated the effect of noise suppression on the loud noise artifact, known as glitch.
We use the Gravity Spy database \cite{GravitySpy2017,GravitySpy2021,GravitySpyZenodo} to obtain various common types of glitches with an estimated SNR larger than 10 and confidence $>$ 0.95.
We focus on three categories (Blip, Scattered Light, and Koi Fish) since they are known to be problematic to mimic the response of detectors to an actual GW event \cite{glitchBBH} and thus limit the overall sensitivity of GW searches \cite{dqligoO1,dqligoO2O3,dqvirgoO3,glitchO3}.
Peak frequency is defined as the frequency with the highest amplitude in the frequency spectrum of the signal.
The bottom panels of Figure ~\ref{fig:glitch} show the comparison of the amplitude at peak frequency between the original and suppressed glitches during the second half of the third observing run (O3b), and its corresponding ASD distribution.
Detailed results of other observing runs (O1, O2 and O3a) are given in \ref{app3}, and they exhibit a similar distribution pattern as O3b.
It can be noticed that the amplitude is compressed to multiple orders of magnitude below its original value.
Take O3b result as example, average compression ratio of Blip, Scattered Light, Koi Fish, and other instances are 78.7, 184.7, 605.7, and 611.4, respectively.
And the ASD distribution is similar as pure noise.
The results indicate that our model can significantly suppress the level of glitch that embedded in real advanced LIGO-Virgo noise.

\subsection{Recovery of binary black holes}\label{subsec2}

\begin{figure*}
    \centering
    \includegraphics[width=1.0\textwidth]{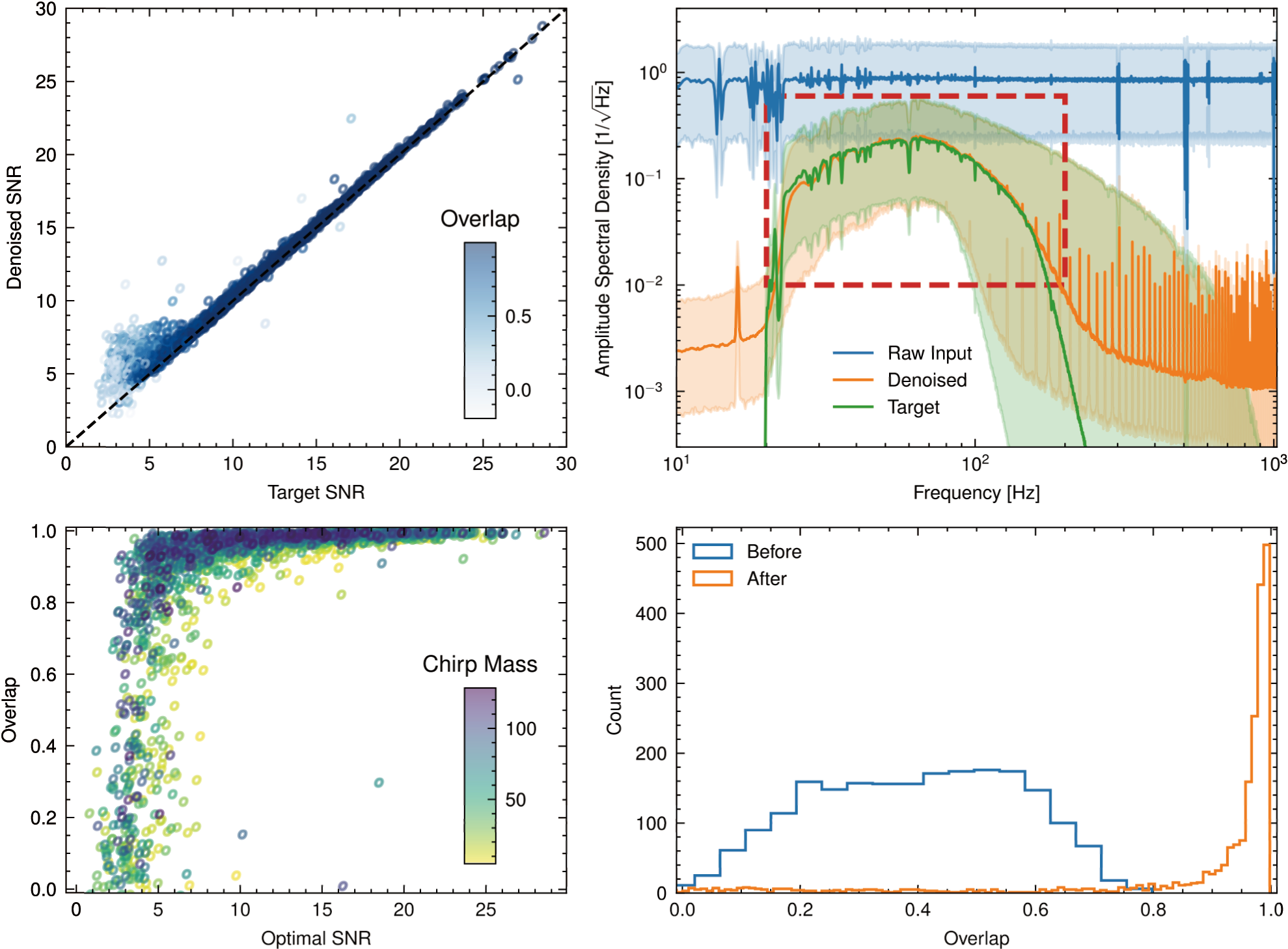}
    \caption{\textbf{Signal recovery results of simulated signal with noise from Hanford O3b data.}
    {\textbf{(upper left)}} Signal amplitude recovery performance between target SNR and denoised data SNR.
    The closer scatter points are to the black dashed line, the better amplitudes are recovered.
    For most waveforms, its corresponding scatter point is quite close to the line.
    {\textbf{(upper right)}} The data ASD after noise suppression.
    Among the frequency region(red dashed rectangle box, 20-200Hz) that contains richest signal information, ASD of our denoised output is significantly consistant with that of target, the relative difference between median of power spectral density distribution is about 7\%.
    {\textbf{(lower left)}} Signal phase recovery performance.
    The overlap is calculated between WaveFormer's denoised output with its corresponding groundtruth whitened waveform to evaluate phase recovery accuracy. Higher optimal SNR and chirp mass both lead to higher overlap.
    {\textbf{(lower right)}} After noise suppression, the overlap is higher than 0.9 for most waveforms, which valids our phase recovery accuracy.
    }
    \label{fig:overlap_snr}
\end{figure*}

Based on pure and loud noise suppression ability, we further validate WaveFormer's signal recovery performance while simultaneously suppressing noise as far as possible.
Specifically, we apply the trained network on BBH injections (see more in Section ~\ref{injtest}) in {LIGO} observation noise and evaluate phase and amplitude recovery accuracy.
Overlap and matched-filtering signal-to-noise (MFSNR) \cite{book2013} are calculated to represent phase and amplitude recovery performance.
We calculate the overlap over the same signal duration \cite{ChatterjeeWen2021} for phase recovery and obtain the similar overlaps with \cite{Bacon2022,Murali2022}.
With respect to the overlap distribution among the validation dataset for {Hanford} O3b (more samiliar results in other observating runs are provided in \ref{app3}), overlap is higher than 0.9 for most waveforms (Figure ~\ref{fig:overlap_snr}d), and as expected, higher SNR leads to better overlap preformance (Figure ~\ref{fig:overlap_snr}c) with injections in LIGO-Virgo noise for all three observations, which is consistent with \cite{ChatterjeeWen2021,Bacon2022}.
When optimal SNR $>$ 6, overlaps of almost all samples, specifically 94.58\%, are higher than 90\%.
We also observe that the WaveFormer is slightly biased against the low-mass systems.
Around overlaps of 13\% samples are smaller than 0.90 when chirp mass $<$ 25 solar masses and optimal SNR $>$ 6.
For high-mass systems, $>$ 96\% samples have overlaps $>$ 90\%.

These results demonstrate that the phase information of GW waveform can be accurately recovered using WaveFormer.
Figure ~\ref{fig:overlap_snr}c also shows a comparison between the injected templates and denoised waveforms using MFSNR.
As expected, the denoised SNR comes quite close to the target one in cases of high overlap.
Lower overlap cases have denoised SNR with larger variance.
The root-mean-square residuals for Hanford O3b is $0.53\pm 0.83$, which is significantly better than the results of \cite{Bacon2022} {for Livingston O1 data}. 
As shown in Figure ~\ref{fig:overlap_snr}b, we go deeper and analyze ASD of WaveFormer's denoised output.
Among the intermediate frequency range (20-200Hz) that covers rich BBH signal information, the ASD distribution of denoised waveform is evidently consistent with that of target signal.
{The comparison of ASDs shows that the denoised waveform’s amplitude is reconstructed with a median error of about 7\% relative to the target signal’s amplitude, further illustrating the effectiveness of our method in recovering the gravitational wave signal.}

\begin{figure*}
    \centering
    \includegraphics[width=1.0\textwidth]{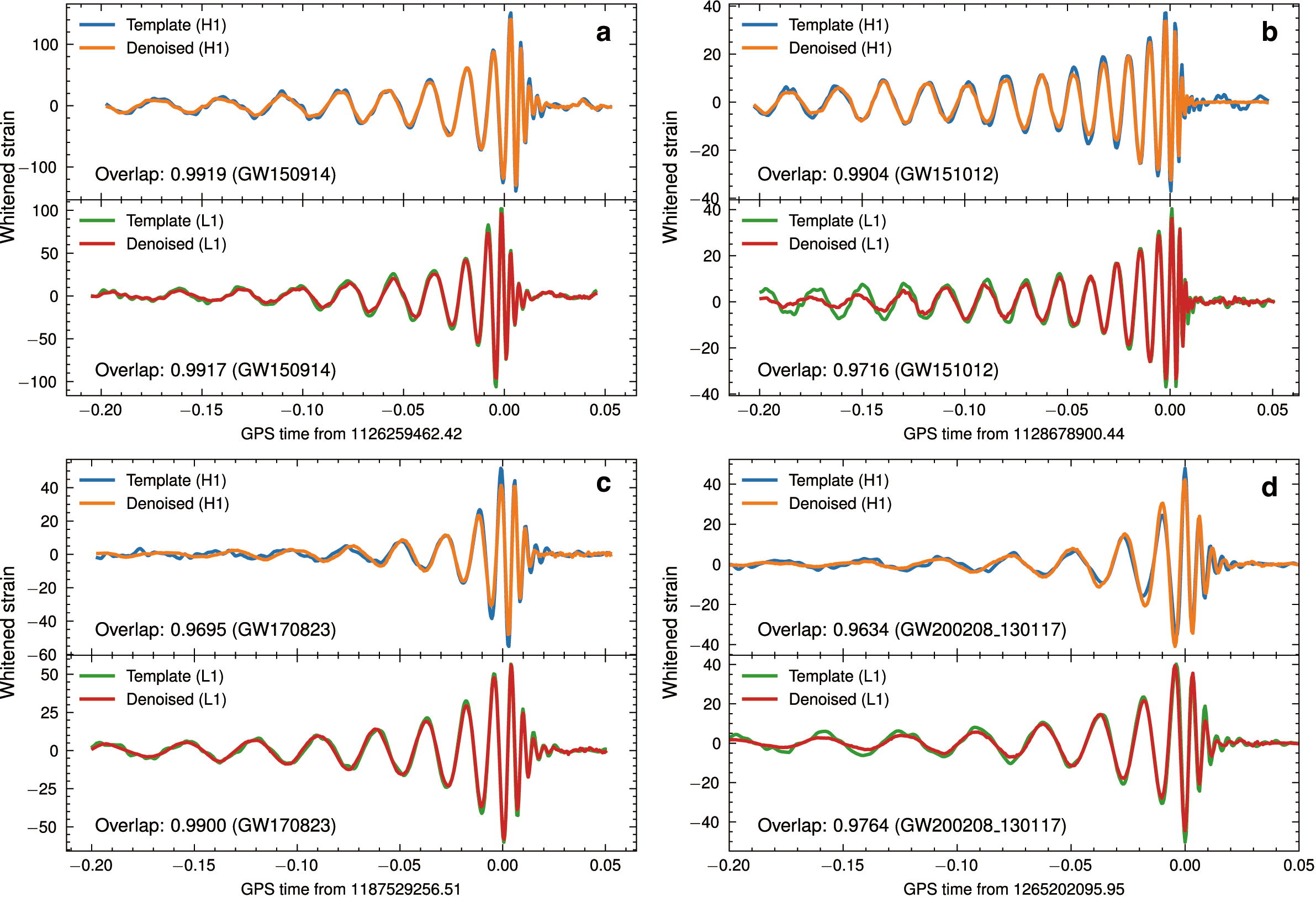}
    \caption{\textbf{Comparison of denoised signals from LIGO observation data for events: a. GW150914, b. GW151012, c. GW170823 and d. GW200208\_130117 with their optimal templates.}
    Hanford and Livingston are represented by H1 and L1.
    For most events, we achieved perfect phase information recovery.
    The ringdown part of GW150914, in particular, can be recovered very well.
    And no matter under what circumstances, like low network SNR (GW151012 and GW200208\_130117), or high chirp mass system (GW170823), the amplitudes of our denoised signal match those of the templates.
    }
    \label{fig:events}
\end{figure*}

Figure ~\ref{fig:events} presents the output of our denoising model when applied to real advanced LIGO noise that contains different BBH events.
We further compare the whitened GW template \cite{GWTC1} with WaveFormer's output and derive four cases from the comparison findings.
Figure ~\ref{fig:events}a shows the first successful detection of the GW signal, GW150914.
We achieved perfect recovery of the inspiral and merger phases at both detectors.
Compared with \cite{WeiHuerta2020,ChatterjeeWen2021}, our result can not only recover the amplitude but also the ringdown part, with an overall overlap $>$ 99.10\% at 0.25-seconds signals around the merger location.
GW151012 has the lowest network SNR, $6.4_{-1.3}^{+1.3}$ for {Hanford} and $5.8_{-1.2}^{+1.2}$ for Livingston, among all BBH events in GWTC-1, hence Bacon et al. \cite{Bacon2022} poorly recovered both the phase and amplitude, while our model shows the ability to retrieve clean cycles.
We completely recovered phase information and obtained the amplitudes reasonably well at mergers and ringdowns of GW151012 (Figure ~\ref{fig:events}b).
The signal overlaps for Handfold and Livingston are 99.04\% and 97.16\%, respectively.
In case of the GW170823 (Figure ~\ref{fig:events}c), a BBH event with high chirp mass $29.2M_\odot$, both Bacon et al. \cite{Bacon2022} and Murali et al. \cite{Murali2022} could recover the phase of original GW signal with certain cycles but failed to recover the complete evaluation in amplitude scale.
In the contrast, we observe a clear match in the amplitude of peaks of the extracted GW170823 waveform, with an overlap of 96.95\% and 99.00\% for Handfold and Livingston, respectively.
Figure ~\ref{fig:events}d shows the most recent detected BBH candidate, GW200208\_130117, during O3b obeservation.
Its network SNR is as low as GW151012, $10.8_{-0.4}^{+0.3}$ to be exact, and we can well recover the GW signal.
These results show that our denoising algorithm outperformed others by capturing the characteristic chirping morphology of BBH evolution, and can denoise signals in realistic detection scenarios without affecting signal characteristics such as phase and amplitude.

\subsection{Significance estimates}\label{subsec4}

\begin{figure*}
    \centering
    \includegraphics[width=1.0\textwidth]{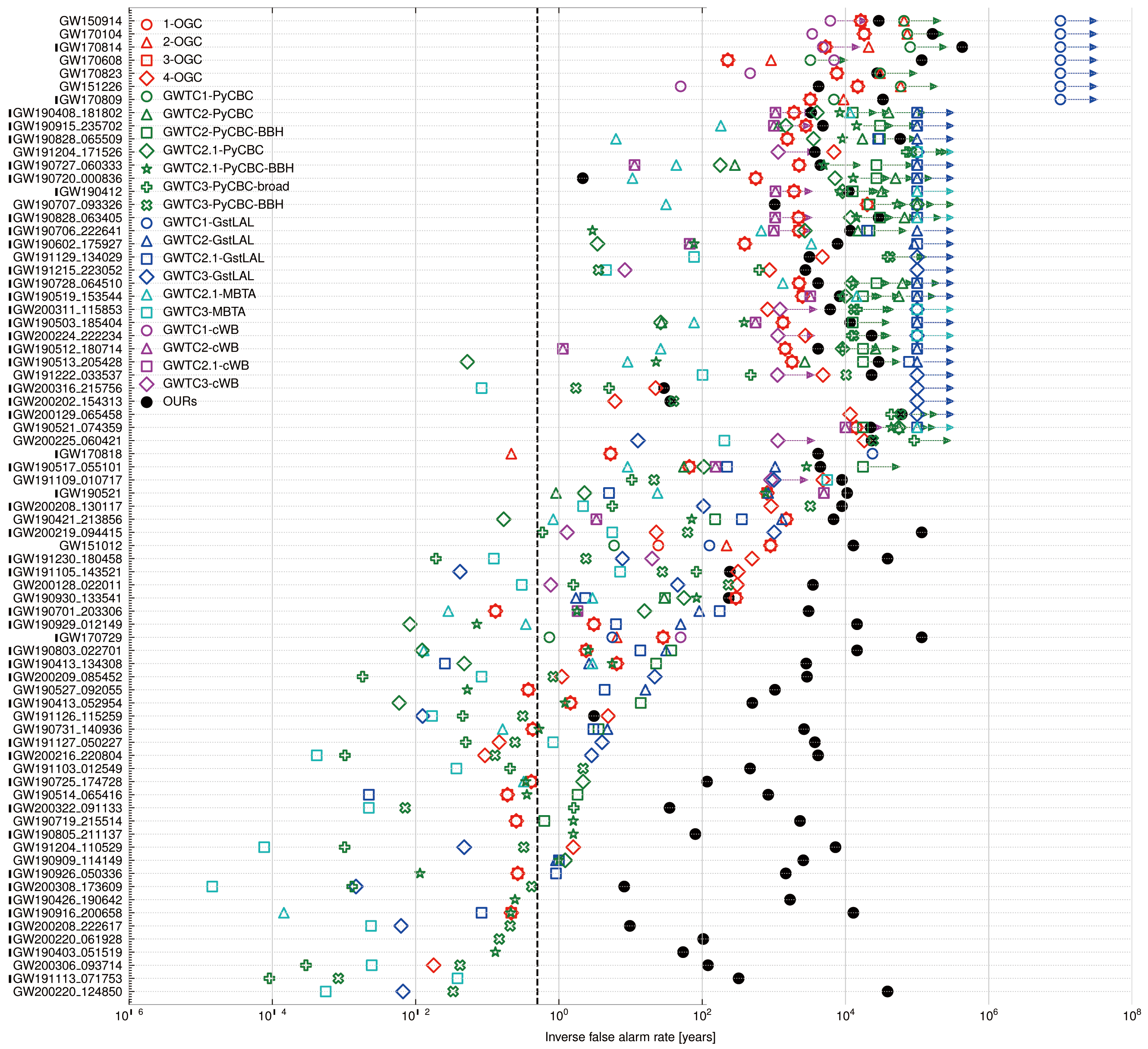}
    \caption{\textbf{IFAR for LIGO BBH events.}
    The BBH events are collected from the first, second, and third Gravitational-Wave Transient Catalogs (GWTC-1/2/2.1/3) \cite{GWTC1,GWTC2,GWTC2.1,GWTC3} and sorted by FAR (from low to high).
    The events marked with a dagger ($\dagger$) contain Virgo data.
    The point with an arrow attached represents the minimum IFAR that can be achieved by the pipelines from the reported catalogs.
    The black dotted line represents a $2.0 \textit{yr}{}^{-1}$ FAR threshold used in GWTC-2/2.1/3 catalogs.
    For all reported BBH events, we achieve {significant} IFAR improvement.
    }
    \label{fig:far}
\end{figure*}

{As shown in Figure ~\ref{fig:far}, we assessed the performance of our denoising workflow by comparing results with the GWTC-1, GWTC-2, GWTC2.1, and GWTC-3 catalogs (referred to as the ’reported catalogs’) as well as their associated data releases.
We prioritized the data obtained directly from these releases, ensuring the most accurate and updated
analysis, rather than relying solely on the summary tables or figures in the publications.}
Comparing with our results, the significance estimates from GWTC \cite{GWTC1,GWTC2,GWTC2.1,GWTC3} and OGC \cite{OGC1,OGC2,OGC3,OGC4} have a more significant divergence in the distribution of the IFAR from the reported catalogs.
With regard to all 75 reported BBH events, we achieve significant IFAR improvement, which indicates that loud terrestrial noise is well suppressed.
For example, in the case of the low network SNR event GW200208\_130117 (as shown in Figure ~\ref{fig:events}d), we obtained an IFAR of 8916 years, when the maximum IFAR of other catalogs is less than 4000 years.

{our analysis indicates that the variability in improvement is closely related to the nature of the noise in the original data. The inherent non-Gaussian, non-stationary characteristics of the noise and the varying strategies of different pipelines in signal recognition contribute to the observed
discrepancies in IFAR improvement.
Furthermore, we found that IFAR performance significantly depends on the extent to which it reduces the non-Gaussian noise near each event. This observation suggests that for events
in which IFAR shows substantial improvement, its misleading non-Gaussian noise is effectively
eliminated. Conversely, for events where IFAR underperforms, the denoised data still retains
non-Gaussian characteristics, which is possibly due to inherent systematic errors of WaveFormer.}


\section{Conclusion}\label{sec3}
Large-scale neural network is a powerful tool that allows us to directly apply machine learning algorithms to raw observational GW data to perform data processing.
We develop an AI-based workflow centered with WaveFormer to achieve accurate and real-time GW noise suppression.
Our proposed WaveFormer model is based on transformer, but with several science-driven innovations.
The combination of convolutional neural network and transformer enables our model the ability of extracting hierarchical features, which correspond to GW signal information of a wide frequency range from science perspective.
Moreover, a masked loss mechanism is proposed and applied.
It can distinguish the recovery importance of different sampling points and assign an appropriate mask accordingly.

All the proposed adaptions have been proven to improve noise suppression performance as well as stabilize and accelerate network convergence.
Firstly, we directly evaluate model's noise suppression on pure noise and glitches.
With regard to pure noise suppresion, noise is significantly supressed. 
Standard deviation of noise amplitude and noise ASD of the whole frequency range are significantly decreased by an order at least.
For different glitch categories, glitch amplitude can be compressed to multiple orders below its original value.
Secondly, we further validate model's signal recovery performance.
On real observational data and BBH events, we achieve state-of-the-art results compared with other deep-learning-based denoising method. WaveFormer can recover the amplitude of low network SNR events and high chirp mass events while other methods fail.
Finally, through significance estimates, we prove that there is a dramatic data quality improvement with our AI-based denoising workflow, and achieve significant IFAR improvement on 75 reported BBH.

To this end, along with the provided applications, this work can be a starting step towards the GW search strategy that can potentially be extended and contributed to the upcoming and heavy data processing and GW search procedures of the fourth observing run.

\section*{Appendix}

\subsection*{Mask definition}\label{app1}

\begin{figure*}
    \centering
    \includegraphics[width=0.6\textwidth]{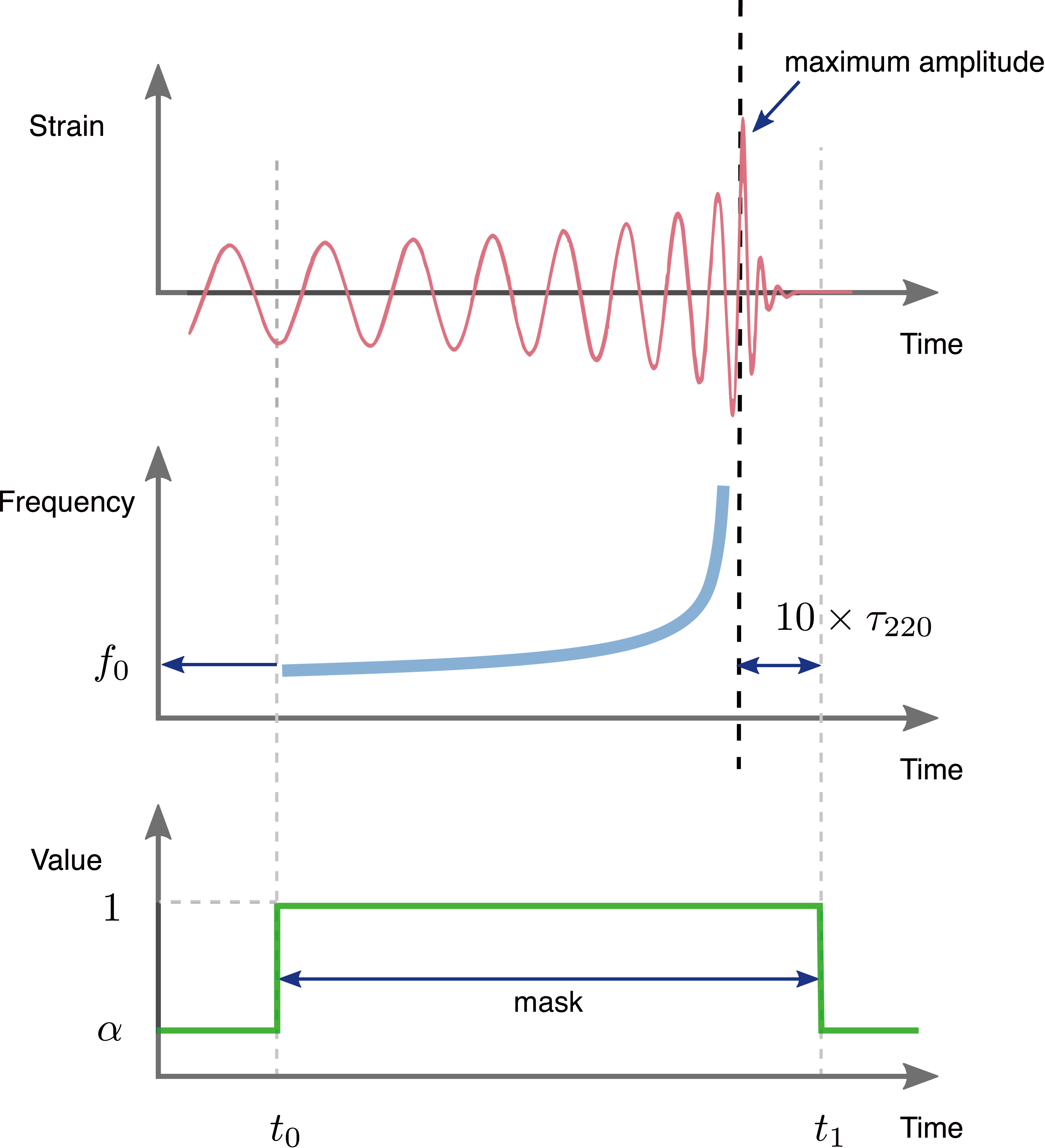}
    \caption{\textbf{Dynamic mask definition strategy.}
    Top panel: An example of a chirp-like time strain with increasing frequency and amplitude, location of the maximum absolute value is considered as the merger moment.
    Middle panel: Time-frequency representation of the above chirp-like signal with low-frequency cutoff at $f_0$.
    $10\times\tau_{220}$ represents ten times of the damping time of dominant quasinormal mode.
    Botton panel: The showcase of mask sequence, based on the characteristic chirping morphology of the above BBH evolution, Values within mask are set to 1 and $\alpha$ for others.
    }
    \label{fig:mask}
\end{figure*}

We implement a dynamic masking operation based on the characteristic of each modeled waveform.
As shown in Figure ~\ref{fig:mask}, we assume the signal waveform is well-modeled and maximum absolute value of the waveform locates at merger time.
Left boarder $t_0$ of the mask depends on the lower frequency cutoff at $f_0=20$Hz based on the post-Newtonian theory.
We approximate the evolution of the pre-merger part as $f(t) = \frac{1}{8 \pi \mathcal{M}^{5 / 8}}\left(\frac{5}{t}\right)^{3 / 8}$, where $\mathcal{M}$ represents chirp mass.
For the right boarder $t_1$, we refer to the damping time $\tau$ from linear perturbation theory to ensure the contribution of ringdown phase.
We specify ten times of the damping time, $10\times\tau_{220}$, of the dominant quasinormal mode to ensure that enough waveforms are enclosed for effective denoising.
{Mask values between $t_0$ and $t_1 + 10\tau_{220}$ are set to 1, otherwise $\alpha$.}
{Value of $\alpha$ is decided by the length ratio of BBH waveform (generally 0.5 to 2 seconds) and model input (8.0625 seconds). We did ablation studies on four settings ($\alpha \in \{1/10, 1/6, 1/4, 1\}$) and found that $\alpha=1/6$ performed best. Hence in our experiment, $\alpha$ is set to $1/6$.}
The dynamic mask is then applied to training loss and stablizes training process.

\subsection*{Detailed suppression results}\label{app3}

\begin{figure*}
    \centering
    \includegraphics[width=1.0\textwidth]{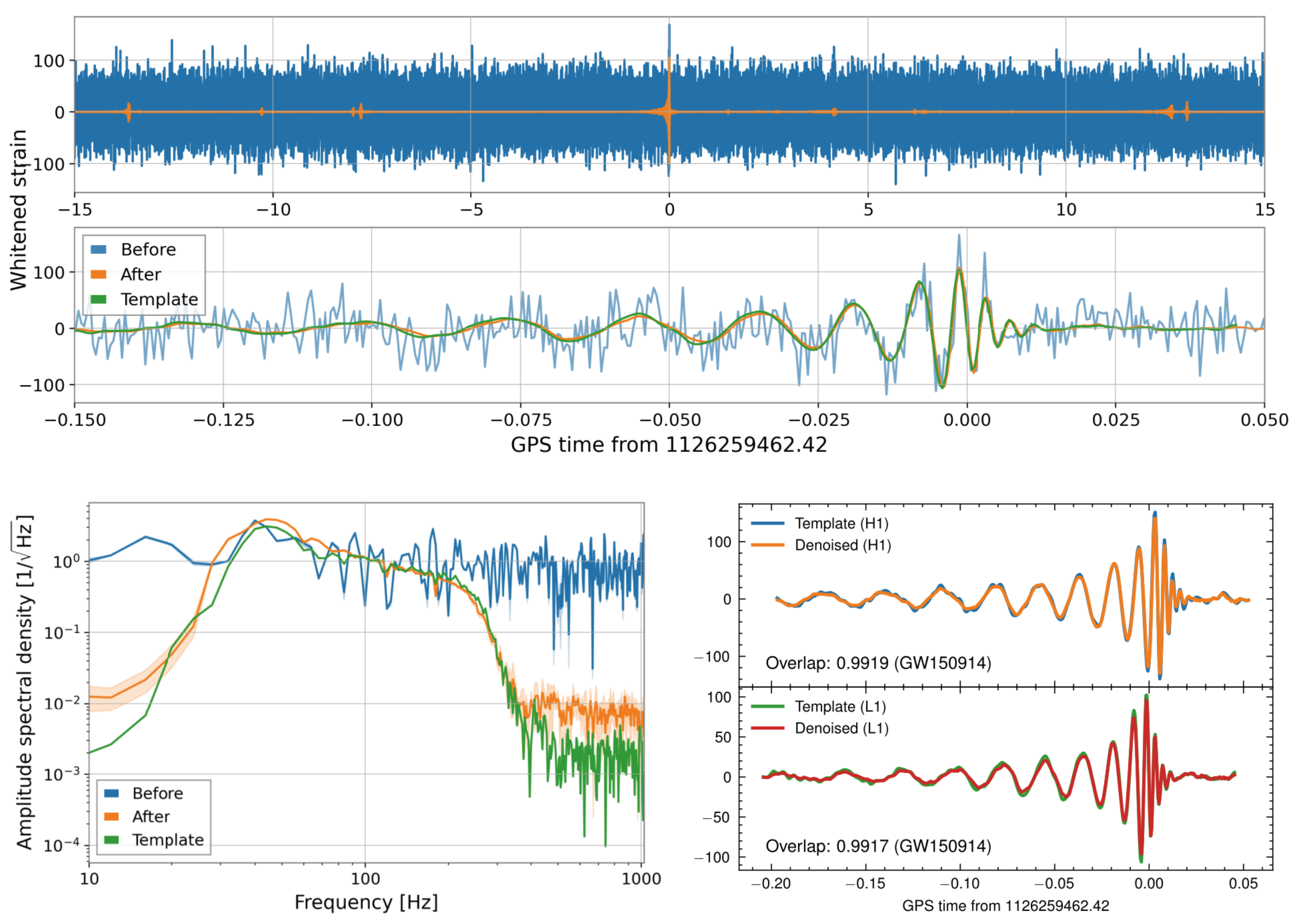}
    \caption{\textbf{Preprocessed input, denoised output, and template of GW150914.}
     a) 30-seconds time-series and 0.2-seconds zoomed-in segment. b) spectrogram. c) 0.25-seconds time-series signal of H1 and L1, respectively.
    }
    \label{fig:sup_signal}
\end{figure*}

\begin{figure*}
    \centering
    \includegraphics[width=1.0\textwidth]{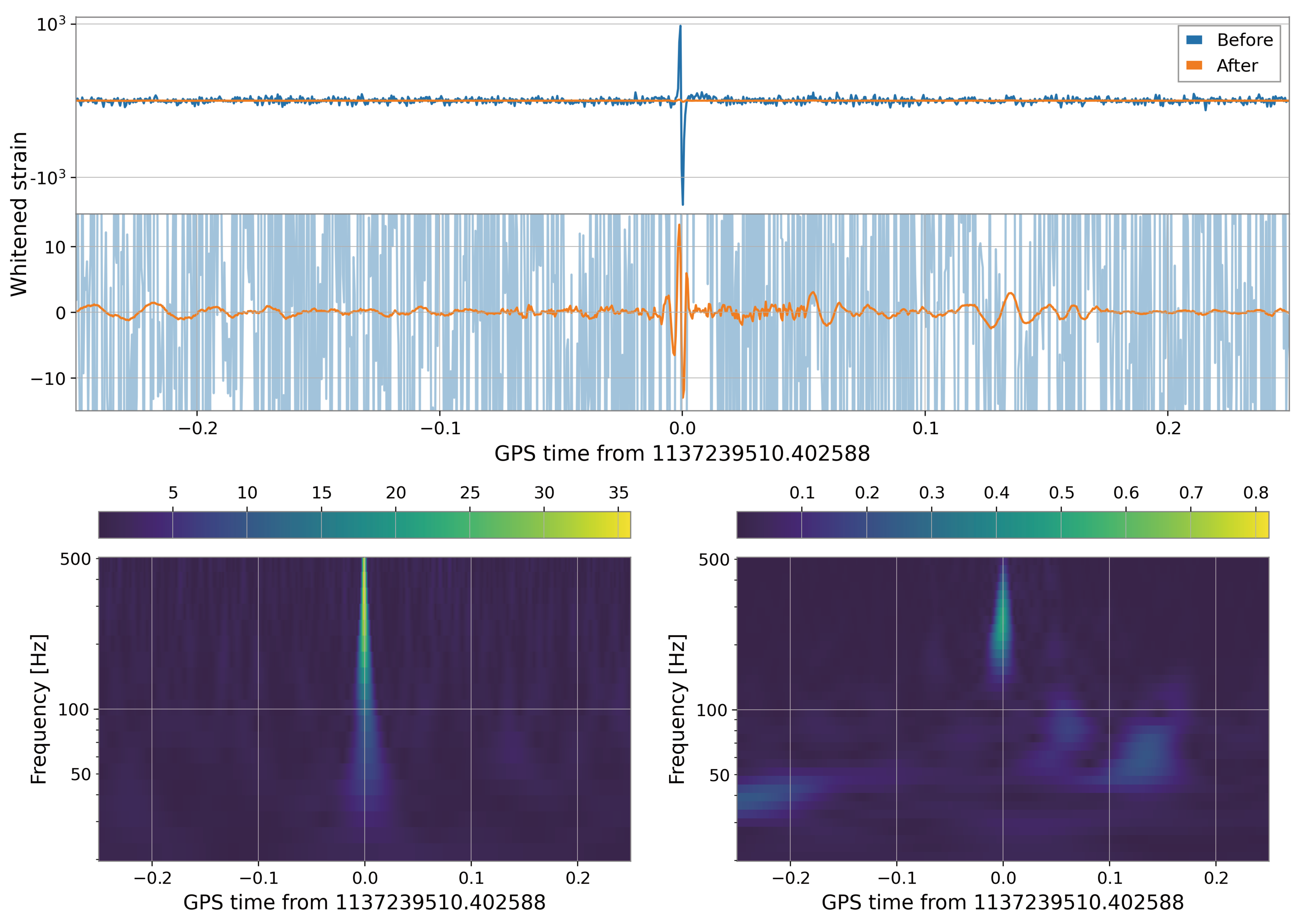}
    \caption{\textbf{Time-series and spectrogram example of blip.}
     a) 30-seconds time-series and 0.2-seconds zoomed-in segment. b) spectrogram of input blip. c) spectrogram of denoised output.
    }
    \label{fig:sup_blip}
\end{figure*}

We provide a clearer picture of the data post-processing, not just from a spectral perspective but also encompassing time series and spectrogram to convey a more holistic view of the method's effectiveness.
{Examples} of signal(GW150914) and blip are shown in Figure \ref{fig:sup_signal} and Figure \ref{fig:sup_blip}.
As for signal, we give a 30-second whole signal and its corresponding 0.2-second zoomed-in segment, as well as spectrogram and denoising performance on both two detectors.
Similarly, the time series and spectrogram of the blip are represented in Figure \ref{fig:sup_blip}.

{Glitch is a common occurrence with a rate of $\lesssim$1 per minute in the LIGO detectors in O3a \cite{GWTC2}.} 
To further investigate the performance of noise suppression on these loud non-Gaussian artifacts, we use the Gravity Spy database \cite{GravitySpy2017,GravitySpy2021,GravitySpyZenodo}, which contains a wide range of glitches.
The total number of LIGO glitches considered in this work from the first three observing runs (O1, O2, and O3, where O3 is divided into O3a and O3b) is 15487, 41497, 101614 and 144958 for O1, O2, O3a, and O3b, respectively.
We set a minimum confidence threshold (0.95) and estimated SNR threshold (10) for all glitch categories to reduce the risk of contamination from the machine learning classifier in Gravity Spy.
In our AI-based workflow, all instances undergo the whitening, normalizing, and WaveFormer denoising preprocessing steps, and the maximum amplitude around each instance's peak frequency is compared to its original value in the whitened domain.

\begin{figure*}
    \centering
    \includegraphics[width=1.0\textwidth]{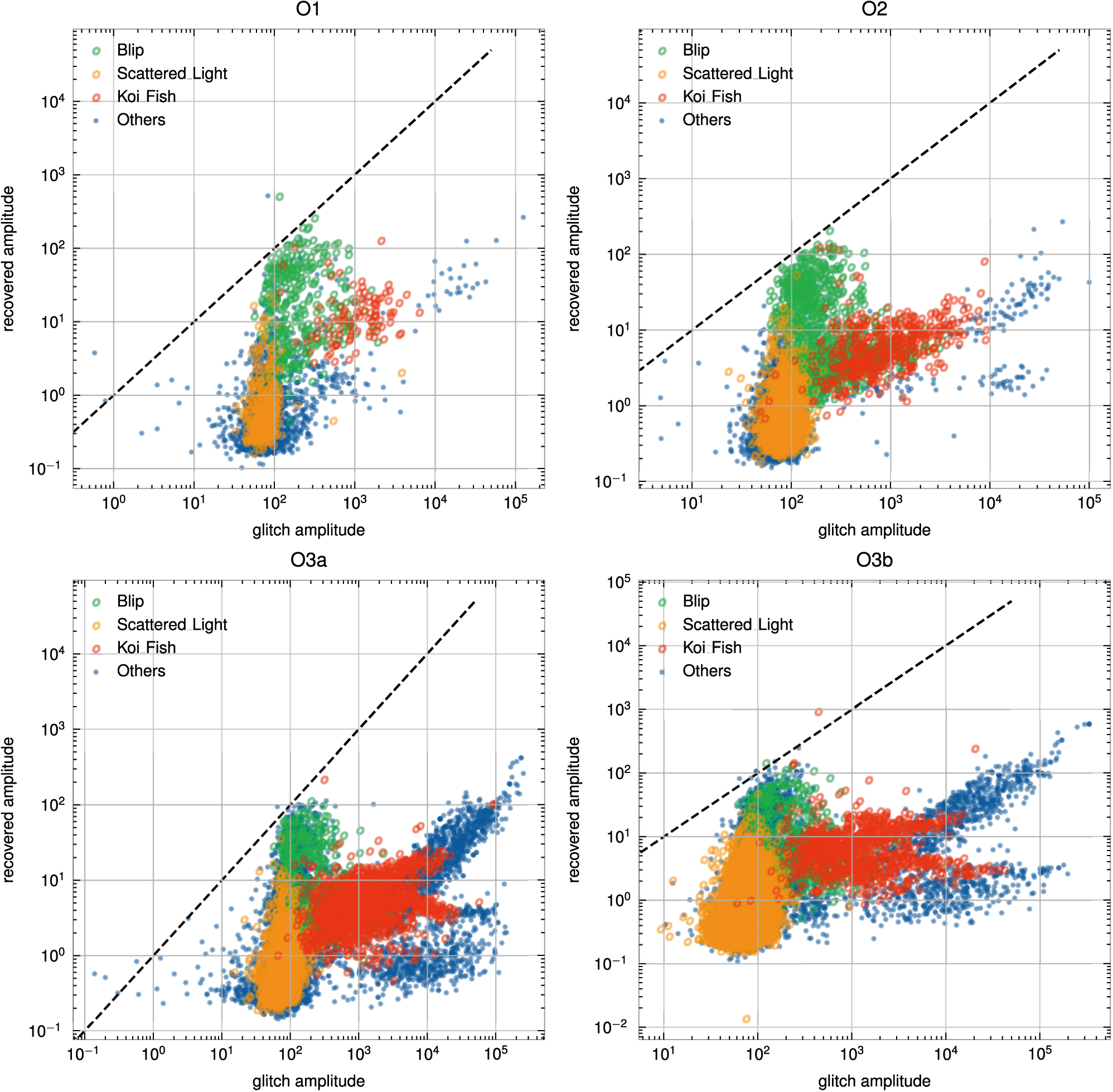}
    \caption{\textbf{Comparison between raw and denoised glitches from the first three observing runs (O1, O2 and O3, where O3 is split into O3a and O3b).}
    The dashed diagonal line represents that the glitch amplitude is equivalent to the denoised glitch amplitude.
    For clarity, 20\% samples of the total are given.
    The four subfigures show similar distribution pattern, and amplitudes of all glitch categories are well suppressed.
    }
    \label{fig:glitch_sup}
\end{figure*}

\begin{figure*}
    \centering
    \includegraphics[width=1.0\textwidth]{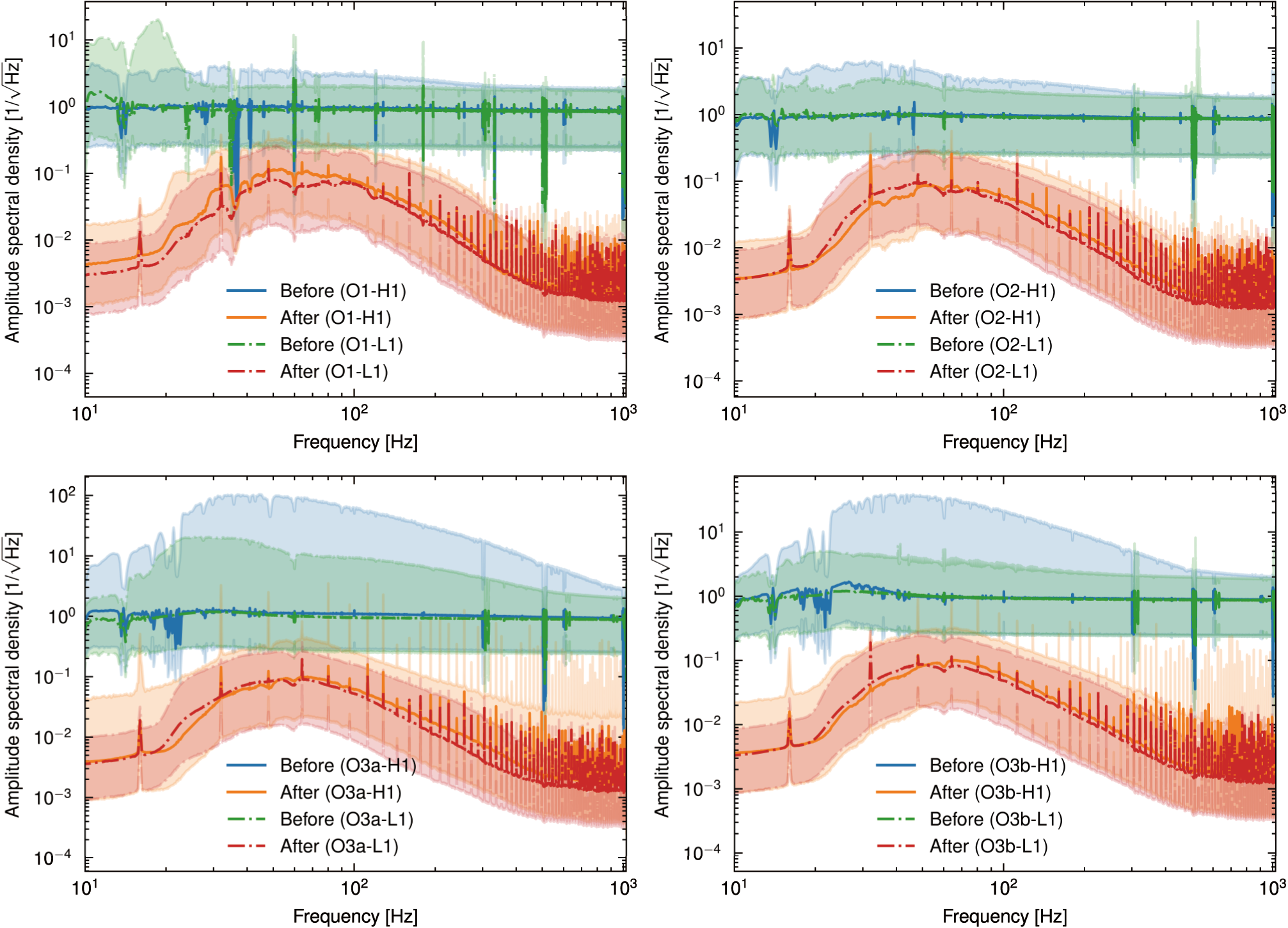}
    \caption{\textbf{The amplitude spectral density comparison between raw and denoised glitches from the first three observing runs (O1, O2 and O3, where O3 is split into O3a and O3b).}
    The noise amplitude spectral density of the whole frequency range is significantly decreased.
    }
    \label{fig:glitch_sup_asd}
\end{figure*}

We focus on three categories (Blip, Scattered Light, and Koi Fish) because they are known to be problematic and can create considerable challenges for candidate event analysis \cite{glitchBBH,dqligoO1,dqligoO2O3,dqvirgoO3,glitchO3}.
Results are given in Figure ~\ref{fig:glitch_sup} and Table \ref{tab:glitch}.
The percentage of instances with higher denoised amplitude ($a^{\textit{denoised}}$) than before ($a^{\textit{original}}$) is quite small and only O1 exceed 1\%.
The average compression ratios, $\frac {\sum_{i=1}^n a^{\textit{original}}_i/a^{\textit{denoised}}_i} {n}$, for each glitch set in O1, O2, O3a, and O3b are all {more than 30 times}.
The noise amplitude spectral density of the glitches from O1, O2, O3a, and O3b is presented in Figure ~\ref{fig:glitch_sup_asd}.

\begin{table}[]
\caption{Noise suppression performance on various glitch categories from the first three observing runs (O1, O2, and O3, where O3 is divided into O3a and O3b).}
\label{tab:glitch}
\begin{tabular}{|r|cccc|}
\hline
\multicolumn{1}{|c|}{} &
  \multicolumn{4}{c|}{\textbf{O1}}
  \\ \hline
\textbf{class} &
  \multicolumn{1}{c|}{\textbf{Blip}} &
  \multicolumn{1}{c|}{\textbf{Scattered Light}} &
  \multicolumn{1}{c|}{\textbf{Koi Fish}} &
  \textbf{others}
  \\ \hline
\begin{tabular}[c]{@{}r@{}}percentage of instances\\ above the diagonal \end{tabular} &
  \multicolumn{1}{c|}{0.0563\%} &
  \multicolumn{1}{c|}{0} &
  \multicolumn{1}{c|}{0} &
  1.7405\%
  \\ \hline
\begin{tabular}[c]{@{}r@{}}average compression ratio \end{tabular} &
  \multicolumn{1}{c|}{34.1661} &
  \multicolumn{1}{c|}{146.5901} &
  \multicolumn{1}{c|}{107.1548} &
  220.9854 \\ \hline
\end{tabular} 

\begin{tabular}{|r|cccc|}
\hline
\multicolumn{1}{|c|}{} &
  \multicolumn{4}{c|}{\textbf{O2}}\\ \hline
\textbf{class} &
  \multicolumn{1}{c|}{\textbf{Blip}} &
  \multicolumn{1}{c|}{\textbf{Scattered Light}} &
  \multicolumn{1}{c|}{\textbf{Koi Fish}} &
  \textbf{others} \\ \hline
\begin{tabular}[c]{@{}r@{}}percentage of instances\\ above the diagonal \end{tabular} &
  \multicolumn{1}{c|}{0.0148\%} &
  \multicolumn{1}{c|}{0} &
  \multicolumn{1}{c|}{0} &
  0.2077\%
  \\ \hline
\begin{tabular}[c]{@{}r@{}}average compression ratio\end{tabular} &
  \multicolumn{1}{c|}{63.3337} &
  \multicolumn{1}{c|}{143.3304} &
  \multicolumn{1}{c|}{216.5543} &
  228.2386 \\ \hline
\end{tabular} 

\begin{tabular}{|r|cccc|}
\hline
\multicolumn{1}{|c|}{} &
  \multicolumn{4}{c|}{\textbf{O3a}} \\ \hline
\textbf{class} &
  \multicolumn{1}{c|}{\textbf{Blip}} &
  \multicolumn{1}{c|}{\textbf{Scattered Light}} &
  \multicolumn{1}{c|}{\textbf{Koi Fish}} &
  \textbf{others} \\ \hline
\begin{tabular}[c]{@{}r@{}}percentage of instances\\ above the diagonal \end{tabular} &
  \multicolumn{1}{c|}{0.0221\%} &
  \multicolumn{1}{c|}{0.0178\%} &
  \multicolumn{1}{c|}{0} &
  0.6837\% \\ \hline
\begin{tabular}[c]{@{}r@{}}average compression ratio\end{tabular} &
  \multicolumn{1}{c|}{92.3541} &
  \multicolumn{1}{c|}{157.2055} &
  \multicolumn{1}{c|}{539.8278} &
  860.3279 \\ \hline
\end{tabular} 

\begin{tabular}{|r|cccc|}
\hline
\multicolumn{1}{|c|}{} &
  \multicolumn{4}{c|}{\textbf{O3b}} \\ \hline
\textbf{class} &
  \multicolumn{1}{c|}{\textbf{Blip}} &
  \multicolumn{1}{c|}{\textbf{Scattered Light}} &
  \multicolumn{1}{c|}{\textbf{Koi Fish}} &
  \textbf{others} \\ \hline
\begin{tabular}[c]{@{}r@{}}percentage of instances\\ above the diagonal \end{tabular} &
  \multicolumn{1}{c|}{0.0900\%} &
  \multicolumn{1}{c|}{0.0075\%} &
  \multicolumn{1}{c|}{0.0216\%} &
  0.0450\% \\ \hline
\begin{tabular}[c]{@{}r@{}}arvange compression ratio\end{tabular} &
  \multicolumn{1}{c|}{78.7399} &
  \multicolumn{1}{c|}{184.7708} &
  \multicolumn{1}{c|}{605.7298} &
  611.4205 \\ \hline
\end{tabular} 
\end{table}\normalsize

\subsection*{Example of denoising result on NSBH event}\label{rev_NSBH}

\begin{figure*}
    \centering
    \includegraphics[width=1.0\textwidth]{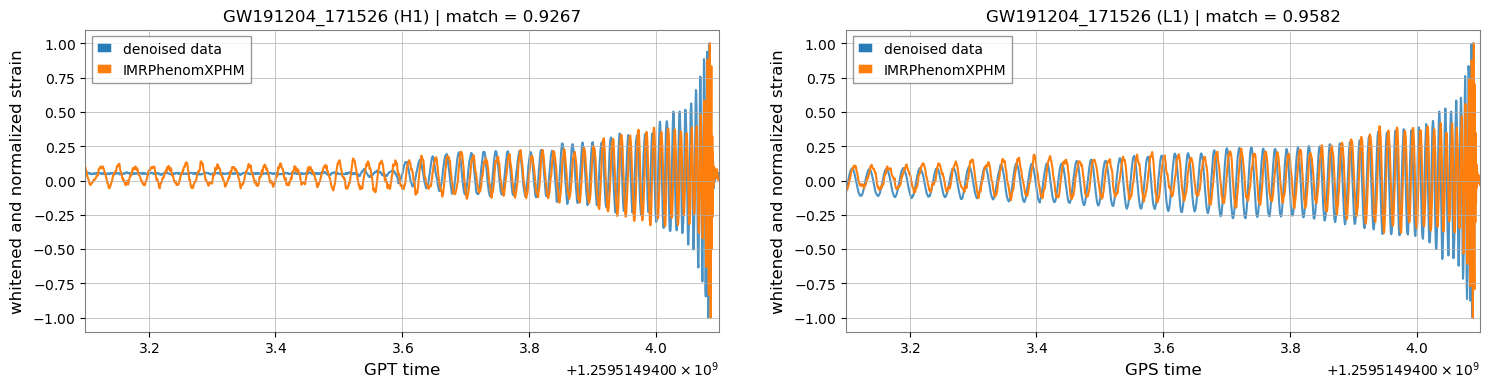}
    \caption{\textbf{WaveFormer's denoising performance on NSBH event (GW191204\_171526).} On H1 and L1 data, we achieve overlap values of 0.9267 and 0.9582 with IMRPhenomXPHM template respectively, which are significantly higher than those of BayesWave and cWB (0.82 to 0.86).
    }
    \label{sup_revfig:GW191204_171526}
\end{figure*}

{From a scientific standpoint, the comprehensive hierarchical feature extraction mechanism can process long signals with up to 8.0625 seconds in length and learn rich gravitational wave information.
This input segment length, while seemingly excessive for typically short BBH signals, is actually crucial for effective noise modeling and signal reconstruction. 
Such a design is an advantage not only for BBH signals but also shows promise for longer signal types such as NSBHs and BNSs.
Hence, we further conducted additional tests to explore the potential
of WaveFormer in the context of NSBH signals.}

{Our preliminary tests on available NSBH data from O1 to O3 observing runs revealed interesting results. Notably, for the event GW191204\_171526, classified as either an NSBH or a low-mass BBH candidate in GWTC-3, WaveFormer demonstrated a significant improvement (Figure \ref{sup_revfig:GW191204_171526}). The overlap with IMRPhenomXPHM achieved 0.93 and 0.95 on H1 and L1, respectively, which are marked improvements over those achieved by BayesWave and cWB (with overlaps between 0.82 to 0.86). This singular yet promising result suggests that while WaveFormer is currently optimized and trained on BBH signals, it holds potential for application to NSBH signals.
It highlights not only WaveFormer's current capabilities but also its potential versatility and adaptability to other types of gravitational wave signals.
The result also demonstrate the significance of WaveFormer as a valuable addition to the IGWN software ecosystem.}

\section*{Code availability}
Our WaveFormer code is avaliable at \href{https://github.com/AI-HPC-Research-Team/LIGO_noise_suppression}{\url{https://github.com/AI-HPC-Research-Team/LIGO_noise_suppression}}.

\section*{Acknowledgments}
This research uses data or software obtained from the Gravitational Wave Open Science Center (gwosc.org), a service of LIGO Laboratory, the LIGO Scientific Collaboration, the Virgo Collaboration, and KAGRA. The research was supported by the Peng Cheng Laboratory and Peng Cheng Cloud-Brain.
This work was supported in part by the National Key Research and Development Program of China (Grant No.~2021YFC2203001 and No.~2020YFC2201501) and in part by the NSFC (No.~11920101003, No.~12021003, No.~12075297 and No.~12235019). Z. Cao was supported by CAS Project for Young Scientists in Basic Research YSBR-006.








\begin{thebibliography}{99}  
\bibitem{aligo} Aasi, Junaid, et al. “Advanced ligo." Classical and quantum gravity 32.7 (2015): 074001.
\bibitem{GW150914} Abbott, Benjamin P., et al. “GW150914: First results from the search for binary black hole coalescence with Advanced LIGO." Physical Review D 93.12 (2016): 122003.
\bibitem{PRL061102} Abbott, Benjamin P., et al. “Observation of gravitational waves from a binary black hole merger." Physical review letters 116.6 (2016): 061102.
\bibitem{PRX041015} Abbott, Benjamin P., et al. “Binary black hole mergers in the first advanced LIGO observing run." Physical Review X 6.4 (2016): 041015.
\bibitem{avirgo} Acernese, Fet al, et al. “Advanced Virgo: a second-generation interferometric gravitational wave detector." Classical and Quantum Gravity 32.2 (2014): 024001.
\bibitem{GWTC1} Abbott, B. P., et al. “GWTC-1: a gravitational-wave transient catalog of compact binary mergers observed by LIGO and Virgo during the first and second observing runs." Physical Review X 9.3 (2019): 031040.
\bibitem{GWTC2} Abbott, R., et al. “GWTC-2: compact binary coalescences observed by LIGO and Virgo during the first half of the third observing run." Physical Review X 11.2 (2021): 021053.
\bibitem{GWTC2.1} Abbott, R., et al. “Gwtc-2.1: Deep extended catalog of compact binary coalescences observed by ligo and virgo during the first half of the third observing run." arXiv preprint arXiv:2108.01045 (2021).
\bibitem{GWTC3} Abbott, R., et al. “GWTC-3: compact binary coalescences observed by LIGO and Virgo during the second part of the third observing run." arXiv preprint arXiv:2111.03606 (2021).
\bibitem{Burst} Klimenko, Sergey, et al. “A coherent method for detection of gravitational wave bursts." Classical and Quantum Gravity 25.11 (2008): 114029.
\bibitem{CBC1} Dax, Maximilian, et al. “Real-time gravitational wave science with neural posterior estimation." Physical review letters 127.24 (2021): 241103.
\bibitem{CBC2} Dax, M., et al. “Neural Importance Sampling for Rapid and Reliable Gravitational-Wave Inference,(2022)." arXiv preprint arXiv:2210.05686.
\bibitem{MF} Sathyaprakash, Bangalore Suryanarayana, and S. V. Dhurandhar. “Choice of filters for the detection of gravitational waves from coalescing binaries." Physical Review D 44.12 (1991): 3819.
\bibitem{glitchBBH} Davis, Derek, Laurel V. White, and Peter R. Saulson. “Utilizing aLIGO glitch classifications to validate gravitational-wave candidates." Classical and Quantum Gravity 37.14 (2020): 145001.
\bibitem{DeepClean} Saleem, Muhammed, et al. “Demonstration of Machine Learning-assisted real-time noise regression in gravitational wave detectors." arXiv preprint arXiv:2306.11366 (2023).
\bibitem{NonSENS} Vajente, Gabriele, et al. “Machine-learning nonstationary noise out of gravitational-wave detectors." Physical Review D 101.4 (2020): 042003.
\bibitem{Bayeswave} Cornish, Neil J., and Tyson B. Littenberg. “Bayeswave: Bayesian inference for gravitational wave bursts and instrument glitches." Classical and Quantum Gravity 32.13 (2015): 135012.
\bibitem{noiseLIGO1} Nuttall, L. K. “Characterizing transient noise in the LIGO detectors." Philosophical Transactions of the Royal Society A: Mathematical, Physical and Engineering Sciences 376.2120 (2018): 20170286.
\bibitem{noiseLIGO2} Berger, Beverly K. “Identification and mitigation of Advanced LIGO noise sources." Journal of Physics: Conference Series. Vol. 957. No. 1. IOP Publishing, 2018.
\bibitem{dqligoO1} Abbott, Benjamin P., et al. “Effects of data quality vetoes on a search for compact binary coalescences in Advanced {LIGO's} first observing run." Classical and Quantum Gravity 35.6 (2018): 065010.
\bibitem{dqligoO2O3} Davis, Derek, et al. “LIGO detector characterization in the second and third observing runs." Classical and Quantum Gravity 38.13 (2021): 135014.
\bibitem{dqvirgoO3} Acernese, F., et al. “Virgo Detector Characterization and Data Quality: results from the O3 run." arXiv preprint arXiv:2210.15633 (2022).
\bibitem{denoiseLIGO} Davis, Derek, et al. “Improving the sensitivity of Advanced LIGO using noise subtraction." Classical and Quantum Gravity 36.5 (2019): 055011.
\bibitem{MLGWreview1} Cuoco, Elena, et al. “Enhancing gravitational-wave science with machine learning." Machine Learning: Science and Technology 2.1 (2020): 011002.
\bibitem{MLGWreview2} Huerta, E. A., and Zhizhen Zhao. “Advances in machine and deep learning for modeling and real-time detection of multi-messenger sources." Handbook of Gravitational Wave Astronomy (2020): 1-27.
\bibitem{MLGWreview3} Cuoco, Elena, et al. “Computational challenges for multimodal astrophysics." Nature Computational Science 2.8 (2022): 479-485.
\bibitem{MLGWSC1} Schafer, Marlin B., et al. “MLGWSC-1: The first Machine Learning Gravitational-Wave Search Mock Data Challenge." arXiv preprint arXiv:2209.11146 (2022).
\bibitem{ShenHuerta2019} Shen H, George D, Huerta E A, et al. Denoising gravitational waves with enhanced deep recurrent denoising auto-encoders[C]//ICASSP 2019-2019 IEEE International Conference on Acoustics, Speech and Signal Processing (ICASSP). IEEE, 2019: 3237-3241.
\bibitem{WeiHuerta2020} Wei W, Huerta E A. Gravitational wave denoising of binary black hole mergers with deep learning[J]. Physics Letters B, 2020, 800: 135081.
\bibitem{ChatterjeeWen2021} Chatterjee, Chayan, et al. “Extraction of binary black hole gravitational wave signals from detector data using deep learning." Physical Review D 104.6 (2021): 064046.
\bibitem{Bacon2022} Bacon, Philippe, Agata Trovato, and M. Bejger. “Denoising gravitational-wave signals from binary black holes with dilated convolutional autoencoder." arXiv preprint arXiv:2205.13513 (2022).
\bibitem{Murali2022} Murali, Chinthak, and David Lumley. “Detecting and Denoising Gravitational Wave Signals from Binary Black Holes using Deep Learning." arXiv preprint arXiv:2210.01718 (2022).
\bibitem{deepclean} Ormiston, Rich, et al. “Noise reduction in gravitational-wave data via deep learning." Physical Review Research 2.3 (2020): 033066.
\bibitem{alphafold2} Jumper, John, et al. “Highly accurate protein structure prediction with AlphaFold." Nature 596.7873 (2021): 583-589.
\bibitem{bert} Devlin, Jacob, et al. “Bert: Pre-training of deep bidirectional transformers for language understanding." arXiv preprint arXiv:1810.04805 (2018).
\bibitem{gpt3} Brown, Tom, et al. “Language models are few-shot learners." Advances in neural information processing systems 33 (2020): 1877-1901.
\bibitem{IGWN1} Bagnasco, Stefano. “Virgo and Gravitational-Wave Computing in Europe." EPJ Web of Conferences. Vol. 245. EDP Sciences, 2020.
\bibitem{IGWN2} Bambi, Cosimo, Stavros Katsanevas, and Konstantinos D. Kokkotas, eds. Handbook of Gravitational Wave Astronomy. Springer Nature, 2022.
\bibitem{IGWN3} Bagnasco, Stefano. “The Ligo-Virgo-KAGRA Computing Infrastructure for Gravitational-wave Research." arXiv preprint arXiv:2311.12559 (2023).
\bibitem{GWOSC1} Vallisneri, Michele, et al. “The LIGO open science center." Journal of Physics: Conference Series. Vol. 610. No. 1. IOP Publishing, 2015.
\bibitem{GWOSC2} Abbott, Rich, et al. “Open data from the first and second observing runs of Advanced LIGO and Advanced Virgo." SoftwareX 13 (2021): 100658.
\bibitem{GWOSC3} Abbott et al. “Open data from the third observing run of LIGO, Virgo, KAGRA and GEO", ApJS 267 29 (2023).
\bibitem{HWinj} Biwer, C., et al. “Validating gravitational-wave detections: The Advanced LIGO hardware injection system." Physical Review D 95.6 (2017): 062002.
\bibitem{IMRPhenomPv21} Khan, Sebastian, et al. “Frequency-domain gravitational waves from nonprecessing black-hole binaries. II. A phenomenological model for the advanced detector era." Physical Review D 93.4 (2016): 044007.
\bibitem{IMRPhenomPv22} Hannam, Mark, et al. “Simple model of complete precessing black-hole-binary gravitational waveforms." Physical review letters 113.15 (2014): 151101.
\bibitem{IMRPhenomPv23} Bohe Alejandro, et al. “Phenompv2 technical notes for lal implementation." LIGO Technical Document, LIGO-T1500602-v4 (2016).
\bibitem{PyCBC} Usman, Samantha A., et al. “The PyCBC search for gravitational waves from compact binary coalescence." Classical and Quantum Gravity 33.21 (2016): 215004.
\bibitem{50overlap} Lazzarini, A., and J. Romano. “Use of overlapping windows in the stochastic background search." LIGO Report, http://www. ligo. caltech. edu/docs (2004).
\bibitem{transformer} Vaswani, Ashish, et al. “Attention is all you need." Advances in neural information processing systems 30 (2017).
\bibitem{palm} Chowdhery, Aakanksha, et al. “Palm: Scaling language modeling with pathways." arXiv preprint arXiv:2204.02311 (2022).
\bibitem{glu} Shazeer, Noam. “Glu variants improve transformer." arXiv preprint arXiv:2002.05202 (2020).
\bibitem{OMFSNR} Eanna E Flanagan, and Scott A. Hughes. “Measuring gravitational waves from binary black hole coalescences. I. Signal to noise for inspiral, merger, and ringdown." Physical Review D 57.8 (1998): 4535.
\bibitem{InnerProd} Finn, Lee S. “Detection, measurement, and gravitational radiation." Physical Review D 46.12 (1992): 5236.
\bibitem{MFSNR} Owen, Benjamin J., and Bangalore Suryanarayana Sathyaprakash. “Matched filtering of gravitational waves from inspiraling compact binaries: Computational cost and template placement." Physical Review D 60.2 (1999): 022002.
\bibitem{OGC1} Nitz, Alexander H., et al. “1-OGC: The first open gravitational-wave catalog of binary mergers from analysis of public Advanced LIGO data." The Astrophysical Journal 872.2 (2019): 195.
\bibitem{OGC2} Nitz, Alexander H., et al. “2-OGC: Open Gravitational-wave Catalog of binary mergers from analysis of public Advanced LIGO and Virgo data." The Astrophysical Journal 891.2 (2020): 123.
\bibitem{OGC3} Nitz, Alexander H., et al. “3-OGC: Catalog of gravitational waves from compact-binary mergers." The Astrophysical Journal 922.1 (2021): 76.
\bibitem{OGC4} Nitz, A. H., et al. “4-OGC: catalogof gravitational waves from compact-binary mergers. arXiv eprints." arXiv preprint arXiv:2112.06878 (2021).
\bibitem{residual} He, Kaiming, et al. “Deep residual learning for image recognition." Proceedings of the IEEE conference on computer vision and pattern recognition. 2016.
\bibitem{megatron} Shoeybi, Mohammad, et al. “Megatron-lm: Training multi-billion parameter language models using model parallelism." arXiv preprint arXiv:1909.08053 (2019).
\bibitem{ray} Moritz, Philipp, et al. “Ray: A distributed framework for emerging {AI} applications." 13th USENIX Symposium on Operating Systems Design and Implementation (OSDI 18). 2018.
\bibitem{pytorch} Paszke, Adam, et al. “Pytorch: An imperative style, high-performance deep learning library." Advances in neural information processing systems 32 (2019).
\bibitem{adam} Kingma, Diederik P., and Jimmy Ba. “Adam: A method for stochastic optimization." arXiv preprint arXiv:1412.6980 (2014).
\bibitem{GravitySpy2017} Zevin, Michael, et al. “Gravity Spy: integrating advanced LIGO detector characterization, machine learning, and citizen science." Classical and quantum gravity 34.6 (2017): 064003.
\bibitem{GravitySpy2021} Soni, Siddharth, et al. “Discovering features in gravitational-wave data through detector characterization, citizen science and machine learning." Classical and Quantum Gravity 38.19 (2021): 195016.
\bibitem{GravitySpyZenodo} Bahaadini, Sara, et al. “Machine Learning for Gravity Spy: Glitch Classification and Dataset". Information Sciences, v1.0.0, vol. 444, Zenodo, 31 Oct. 2018, pp.172¨C186, doi:10.5281/zenodo.1476156.
\bibitem{glitchO3} Davis, Derek, et al. “Subtracting glitches from gravitational-wave detector data during the third observing run." Classical and Quantum Gravity (2022).
\bibitem{book2013} Helstrom, Carl W. Statistical theory of signal detection: international series of monographs in electronics and instrumentation. Vol. 9. Elsevier, 2013.

\end{thebibliography}
\end{document}